\documentclass[twocolumn,times,tighten]{aastex61}

\usepackage{mathptmx}

\usepackage{natbib}
\usepackage{amssymb}

\begin{document}

\title{Spectral and Timing Analysis of the accretion-powered pulsar
4U\,1626$-$67 observed with \textit{Suzaku} and \textit{NuSTAR}}

\author{Wataru B. Iwakiri}\altaffiliation{RIKEN MAXI team, 2-1 Hirosawa, Wako, Saitama 351-0198, Japan}
\affiliation{Department of Physics, Faculty of Science and Engineering, Chuo University, 1-13-27 Kasuga, Bunkyo-ku, Tokyo 112-8551, Japan}

\author{Katja Pottschmidt}
\affiliation{CRESST, Department of Physics, and Center for Space
  Science and Technology, UMBC, Baltimore, MD 21250, USA; NASA Goddard
  Space Flight Center, Greenbelt, MD 20771, USA}

\author{Sebastian Falkner}
\affiliation{Dr.~Karl-Remeis-Sternwarte and ECAP, Sternwartstr.~7,
  96049 Bamberg, Germany}

  \author[0000-0002-1676-6954]{Paul B. Hemphill}
\affiliation{MIT Kavli Institute for Astrophysics and Space Research, Massachusetts Institute of Technology, Cambridge, MA 02139, USA}

\author{Felix F\"{u}rst}\altaffiliation{Cahill Center for Astronomy
  and Astrophysics, California
  Institute of Technology, Pasadena, CA 91125, USA}
\affiliation{European Space Astronomy Centre (ESA/ESAC), Science Operations Department, Villanueva de la Ca\~{n}ada (Madrid), Spain}

\author{Osamu Nishimura}
\affiliation{Department of Electronics and Computer Science, Nagano
  National College of Technology, Nagano 381-8550, Japan}

\author{Fritz-Walter Schwarm}
\affiliation{Dr.~Karl-Remeis-Sternwarte and ECAP, Sternwartstr.~7,
  96049 Bamberg, Germany}

\author{Michael T. Wolff}
\affiliation{Space Science Division, Naval Research Laboratory,
  Washington, DC 20375-5352, USA}

\author{Diana M. Marcu-Cheatham}
\affiliation{CRESST, Department of Physics, and Center for Space
  Science and Technology, UMBC, Baltimore, MD 21250, USA; NASA Goddard
  Space Flight Center, Greenbelt, MD 20771, USA}

\author{Deepto Chakrabarty}
\affiliation{MIT Kavli Institute for Astrophysics and Space Research,
  Massachusetts Institute of Technology, Cambridge, MA 02139, USA}

\author{John A. Tomsick}
\affiliation{Space Sciences Laboratory, University of California,
  Berkeley, CA 94720, USA}

\author{Colleen A. Wilson-Hodge}
\affiliation{ZP 12 Astrophysics Office, NASA Marshall Space Flight
  Center, Huntsville, AL 35812, USA}

\author{Matthias K\"{u}hnel}
\affiliation{Dr.~Karl-Remeis-Sternwarte and ECAP, Sternwartstr.~7,
  96049 Bamberg, Germany}

\author{Yukikatsu Terada}
\affiliation{Graduate School of Science and Engineering, Saitama
  University, 255 Shimo-Okubo, Sakura, Saitama 338-8570, Japan}

\author{Teruaki Enoto}
\affiliation{The Hakubi Center for Advanced Research, Kyoto
  University, Kyoto 606-8302, Japan}

\author[0000-0003-2065-5410]{J\"{o}rn Wilms}
\affiliation{Dr.~Karl-Remeis-Sternwarte and ECAP, Sternwartstr.~7,
  96049 Bamberg, Germany}

\begin{abstract}
We present an analysis of the spectral shape and pulse profile of the
accretion-powered pulsar 4U\,1626$-$67
observed with \textit{Suzaku} and \textit{NuSTAR} during a spin-up state.
The pulsar, which experienced a torque reversal to spin-up in 2008,
has a spin period of $\sim$7.7\,s. Comparing the phase-averaged
spectra obtained with \textit{Suzaku} in 2010 and with \textit{NuSTAR}
in 2015, we find that the spectral shape changed between the two
observations: the 3--10\,keV flux increased by $\sim$5\% while the
30--60\,keV flux decreased significantly by $\sim$35\%. Phase-averaged
and phase-resolved spectral analysis shows that the continuum spectrum
observed by \textit{NuSTAR} is well described by an empirical NPEX
continuum with an added broad Gaussian emission component around the
spectral peak at $\sim$20\,keV. Taken together with the observed
$\dot{P}$ value obtained from \textit{Fermi}/GBM, we conclude that the
spectral change between the \textit{Suzaku} and \textit{NuSTAR}
observations was likely caused by an increase of the accretion rate.
We also report the possible detection of asymmetry in the profile of
the fundamental cyclotron line. Furthermore, we present a study of the
energy-resolved pulse profiles using a new relativistic ray tracing
code, where we perform a simultaneous fit to the pulse profiles
assuming a two-column geometry with a mixed pencil- and fan-beam
emission pattern. The resulting pulse profile decompositions enable us
to obtain geometrical parameters of accretion columns (inclination, azimuthal and polar angles) and a fiducial set of beam patterns. This information is important to validate the theoretical predictions from radiation transfer in a strong magnetic field. 
\end{abstract}


\section{Introduction}\label{sec:intro}

The magnetic field strengths of neutron stars
can be measured directly by observing cyclotron resonance scattering
features (CRSFs, or cyclotron lines) in their hard X-ray spectra. 
Since the parameters of observed CRSFs are determined by the properties of the
accreted plasma \citep[e.g.,][]{mes92,sch07,nishimura08,sch17a,sch17b}, CRSFs
in principle provide us with powerful probes of physical processes in strong
magnetic fields. This is an active field of research.

The first CRSF was discovered in the X-ray spectrum of Her X-1 by
\citet{tru78}. To date, CRSFs have been detected from over 20 sources
\citep[e.g.,][]{miha95,cab12,tom15}, with observed magnetic field
strengths ranging from $10^{12}$\,G to $10^{13}$\,G. The
\textit{Nuclear Spectroscopic Telescope Array}
\citep[\textit{NuSTAR};][]{fiona2013} is an ideal tool to study
cyclotron lines, due to its excellent energy resolution and
uninterrupted coverage in the energy band relevant for CRSF
discoveries \citep[e.g.,][]{furst14,ten14,bha15,tsy16,bod16,jai16}.
On the other hand, modern theoretical models for cyclotron lines are still not fully in agreement with observations.
Theory mostly predicts complex line shapes, with simulated CRSFs showing
emission wings and asymmetric profiles. Most observations, however, find CRSFs
which are well approximated by smooth, symmetric profiles. For example,
Her~X-1, one of the brightest CRSF sources on the sky, shows no sign of
asymmetry or emission wings in its CRSF profile, as shown by \citet{furst13}
using \textit{NuSTAR} and \textit{Suzaku} data. Legitimate cases of asymmetric
CRSFs are rare in the observational literature, with the most notable example
probably being Cep~X-4, which shows extra absorption in its red wing
\citep{furst15}. There is also the case of V\,0332+53, which may have an
asymmetric profile \citep{katja05}, although \textit{NuSTAR} data show that the
significance of the asymmetric profile depends on the continuum model
\citep{doroshenko17}.

A notable candidate for a complex CRSF profile is 4U\,1626$-$67. This
ultracompact X-ray binary \citep[orbital period 42\,min; see][]{mid81,cha98}
hosts a 7.7\,sec pulsar with a cyclotron line at $\sim$37\,keV
\citep{orl98,cob02,iwa12,cam12}. The pulsar has undergone two torque reversals
in recorded history, in 1990 \citep{cha97} and 2008 \citep{cam10}, and is
currently spinning up. \textit{Suzaku} observations bracketing the torque
reversal found no changes in the CRSF parameters, despite a factor of $\sim$2.8
increase in X-ray flux and large changes in the soft X-ray spectrum
\citep{cam12}. However, a pulse-phase-resolved study of the 2006
\textit{Suzaku} observation (during the spin-down state) by \citet{iwa12}
reported the possible detection of an emission-line-like feature at the CRSF
energy in the dim phase of the pulse profile. The statistics of the 2010
\textit{Suzaku} observation were too limited to study the CRSF profile in great
detail, despite the higher flux.

An analysis of the \textit{NuSTAR} observation of 2015 May by
\citet[][hereafter D17]{dai17} also showed the CRSF to be asymmetric.
D17 model the phase-averaged broadband spectrum obtained by
\textit{NuSTAR} and \textit{Swift} with the bulk+thermal
Comptonization continuum model of \citet{becker07} and an additional
component modeled as disk reflection \citep{ballantyne12} with two
CRSFs. The profile of the first $37.95 \pm0.15$\,keV CRSF is suggested
to be complex, and, in contrast to the earlier \textit{Suzaku}
analysis \citep{cam12,iwa12}, a second harmonic at $61.0 \pm1.0$\,keV
is claimed.

In this paper we re-analyze the \textit{NuSTAR} and \textit{Suzaku}
data from 4U\,1626$-$67\ in order to evaluate the significance of the
asymmetric line profile of the fundamental CRSF and to perform an
analysis of the source behavior using model independent comparisons of the continuum. In addition, to evaluate the geometry
of the accretion column, we perform pulse profile modeling using a new
relativistic ray tracing code. Comparing the derived geometrical properties and beam patterns with previous theoretical works, we can achieve an understanding of the physical processes in the strong magnetic field. The \textit{NuSTAR} and \textit{Suzaku}
observations and data reductions are introduced in
Section~\ref{sec:nustarobs} and~\ref{sec:suzobs}, respectively. We
describe the pulsar's long-term variability in
Section~\ref{sec:varstud}. In Section~\ref{sec:avspec} we model the
phase-averaged X-ray spectra using empirical continua and CRSF models.
Section~\ref{sec:pulseprofile} presents our study of the pulse profile
using a new relativistic ray tracing code. Motivated by the pulse
profile modeling results, Section~\ref{sec:phasespec} presents a
phase-resolved spectral analysis of the \textit{NuSTAR} spectrum. In
Section~\ref{sec:discuss_spectral_var} we discuss the implications and
origins of the observed spectral and timing changes between the
\textit{NuSTAR} and \textit{Suzaku} observations.
Section~\ref{sec:discuss_continuum_emission} investigates the origins
of the observed continuum emission. In Section~\ref{sec:discuss_pulseprofile} we explain the physical interpretation of the pulse profile modeling results comparing with previous theoretical results which take into account the anisotropy of the Thomson scattering cross section in a strong magnetized plasma. Finally,
Section~\ref{sec:discuss_crsfprofile} presents a comprehensive look at
the profile of the fundamental CRSF comparing between observations and
between pulse phase intervals.

All spectral analysis in this work was performed with XSPEC, v.12.9.0. Unless
stated otherwise, all error bars are at the 90\% level for one parameter of
interest.

\section{Observations and Data Reduction} \label{sec:observations}

\subsection{\textit{NuSTAR} observation of 4U\,1626$-$67}\label{sec:nustarobs}

\textit{NuSTAR} is NASA's 11th small explorer (SMEX) mission
\citep{fiona2013}. The satellite covers an energy range of 3--79\,keV
with two CdZnTe detectors, Focal Plane Module A and B (FPMA and
FPMB), located at the focal planes of its hard X-ray mirrors.
\textit{NuSTAR} observed 4U\,1626$-$67 from 2015 May 4, 12:46 UT to
May 5, 20:27 UT (MJD\,57146.5319 to 57147.8521). We reduced the data
with version 1.6.0 of the \texttt{nupipeline} software as distributed
with HEASOFT 6.19. After standard screening of the data with v20170503
of the \textit{NuSTAR} calibration data, the net exposure time of the
observation was 65\,ks. A barycentric correction was applied to the
arrival time of each event in the FPMA and FPMB with the
\texttt{barycorr} tool of HEASOFT. We extracted source spectra from
FPMA and FPMB using a circle of radius $100''$ centered on the source.

For the timing analysis, the background light curves were extracted
from a circular region with a radius of $100''$ at the corner of the
\textit{NuSTAR} field of view opposite to the source. For the spectral
analysis, we modeled background spectra applying the \textit{NuSTAR}
background-fitting and -modeling tool, \texttt{nuskybgd}
\citep{wik2014}\footnote{\url{https://github.com/NuSTAR/nuskybgd}}, using
three blank-sky spectra for each telescope extracted from annular regions with
radii $300''$--$400''$, $400''$--$500''$, and~$500''$--$600''$.

\subsection{\textit{Suzaku} observation of 4U\,1626$-$67}\label{sec:suzobs}

\textit{Suzaku} was the fifth Japanese X-ray satellite
\citep{mitsu07}. It was equipped with two types of instruments, the
X-ray Imaging Spectrometer \citep[XIS][]{koya07} and the Hard X-ray
Detector \citep[HXD][]{taka07}. The XIS was a set of four
charge-coupled device (CCD) cameras at the foci of four X-ray
telescopes \citep{ser07}, covering the lower energy band of
0.2--12\,keV. XIS 0, XIS 2, and XIS 3 were front-illuminated devices,
while XIS 1 was back-illuminated. XIS 2 was taken offline after
micrometeorite strikes prior to these observations, so we use XIS 0
and XIS 3 (combined: XIS-FI) as well as XIS 1 (XIS-BI) in this
analysis. The HXD consisted of PIN silicon diodes (HXD-PIN) and $\mathrm{Gd}_2\mathrm{Si}\mathrm{O}_5\mathrm{Ce}$ (GSO) crystal scintillators,
and covered the 10--600\,keV energy band.

The data used here are from the second \textit{Suzaku} observation of
4U\,1626$-$67, which was performed between 2010 September 6, 12:59
UT and 2010 September 7, 05:42 UT (MJD 55445.5410 to 55446.2375). The
first observation of the source had been conducted in 2006 March and
was analyzed by \citet{iwa12}. The XIS was operated with the 1/4
window option and standard clocking, which has a time resolution of
2\,s. As this is comparable to the 7.7\,s pulse period of of 4U
1626$-$67, phase-resolved analysis of the \textit{Suzaku} data is not
practical. Therefore, we used the \textit{Suzaku} data for the
phase-averaged analysis only. We reprocessed and screened the XIS and
HXD data with the standard criteria, using the \textit{Suzaku}
reprocessing tool \texttt{aepipeline} in the HEASOFT v6.19 package
with calibration versions hxd-20110913, xis-20151005, and
xrt-20110630. The XIS spectra and light curves were extracted from a
circular region with a radius of $4.3'$ centered at the source. We
accumulated the XIS background spectra from a source-free region. We
checked the pile-up effect in the same way as in \citet{yama12} and
found that the pile-up fraction was 3\% and 1\% at 5.7 and
28.7\,pixels from the center of the image, respectively. We excluded
the regions which show $>1$\% pile-up fraction so that the pile-up
effect is negligible for the subsequent analysis. In this paper, we
only used the XIS data down to 3\,keV to adjust the energy range of
\textit{NuSTAR}.

Since the HXD was not capable of imaging, we applied the simulated
Non-X-ray Background (NXB) model provided by the \textit{Suzaku} HXD
team \citep{fuka09} and a Cosmic X-ray Background (CXB), assumed to be
the same as the typical model obtained by \textit{HEAO-1}
\citep{bol87}. The net exposures were 20.1\,ks for the XIS-0,1,3 and
18.7\,ks for the HXD, respectively. Different screening criteria for South
Atlantic Anomaly (SAA) passages and Earth elevation angles result in
different exposures for the XIS and the HXD.

\begin{figure}
\includegraphics[width=8.0cm]{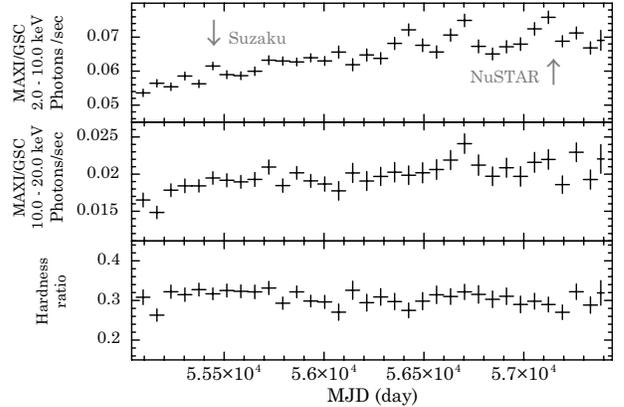}
\caption{Long-term energy-resolved light curves of 4U\,1626$-$67
  obtained with MAXI/GSC. Error bars are at the $1\sigma$
  level. Arrows in the upper panel indicate the dates of the
  \textit{Suzaku} and \textit{NuSTAR} observations.} \label{fig1}
\end{figure}

\subsection{Long-term trend of the X-ray flux below 20\,keV}\label{sec:variab}

Figure~\ref{fig1} shows the long-term light curves and hardness ratio
of 4U\,1626$-$67 from 2009 October to 2016 January, obtained with MAXI
\citep{matsu09} on-board the International Space Station
(\textit{ISS}). To avoid any systematic modulation with a period of
70\,days caused by the \textit{ISS} orbit, the data are binned to a
resolution of 70\,days per bin. Over the period considered here the
X-ray fluxes monotonically increased by $\sim$20\% in both the 2--10
and 10--20\,keV bands, while the hardness ratio remained constant.

\section{Analysis and Results} \label{sec:analysis}

\subsection{Timing analysis}\label{sec:varstud}

Applying epoch folding \citep{leahy83} to the light curves obtained with the
\textit{Suzaku} HXD-PIN and \textit{NuSTAR}, we find the spin period of
4U\,1626$-$67 to be $P_\mathit{Suzaku}=7.6774(1)$\,s and
$P_\mathit{NuSTAR}=7.67295(1)$\,s, respectively. These results are consistent with \textit{Fermi}/GBM
results\footnote{\url{https://gammaray.nsstc.nasa.gov/gbm/science/pulsars.html}}.

We estimate the period derivatives, $\dot{P}$, at the epochs of the
\textit{Suzaku} and \textit{NuSTAR} observations by performing a
linear fit to the \textit{Fermi}/GBM data for 60\,days before and
after each observation. The results are $\dot{P}_\mathit{Suzaku} =
-2.8\times 10^{-11}\,\mathrm{s}\,\mathrm{s}^{-1}$ and
$\dot{P}_\mathit{NuSTAR} = -3.3\times
10^{-11}\,\mathrm{s}\,\mathrm{s}^{-1}$, respectively. Using the
conservative approach of \citet{takagi16}, the uncertainty of these
$\dot{P}$ values is $0.2\times 10^{-11}\,\mathrm{s}\,\mathrm{s}^{-1}$.
Thus, the period derivative $\dot{P}$, along with the X-ray flux
(Figure~\ref{fig1}), increased significantly between the two epochs of
the \textit{Suzaku} and \textit{NuSTAR} observations. Accretion torque
theory \citep{GL79} implies that this change in $\dot{P}$ is due to an
increase in the accretion rate on the neutron star, consistent with
the study of \citet{takagi16}, who find from observations of 4U
1626$-$67 spanning 30\,years that the period derivative changes are in
good agreement with the prediction by \citet{GL79}.

\begin{figure}\centering
\includegraphics[width=6.0cm,angle=270]{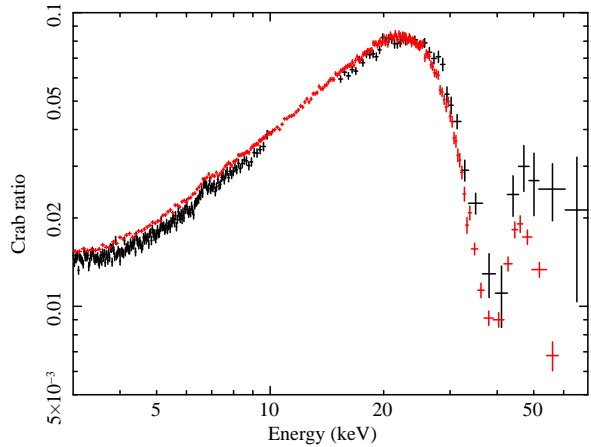}
\caption{Phase-averaged energy-spectra of 4U\,1626$-$67 divided by the
  Crab spectrum. Black and red crosses show the data obtained with
  \textit{Suzaku} (XIS-FI and HXD-PIN) and \textit{NuSTAR} (FPMA),
  respectively. }
\label{fig4}
\end{figure}

\begin{figure*}
\centering
\includegraphics[width=14.0cm]{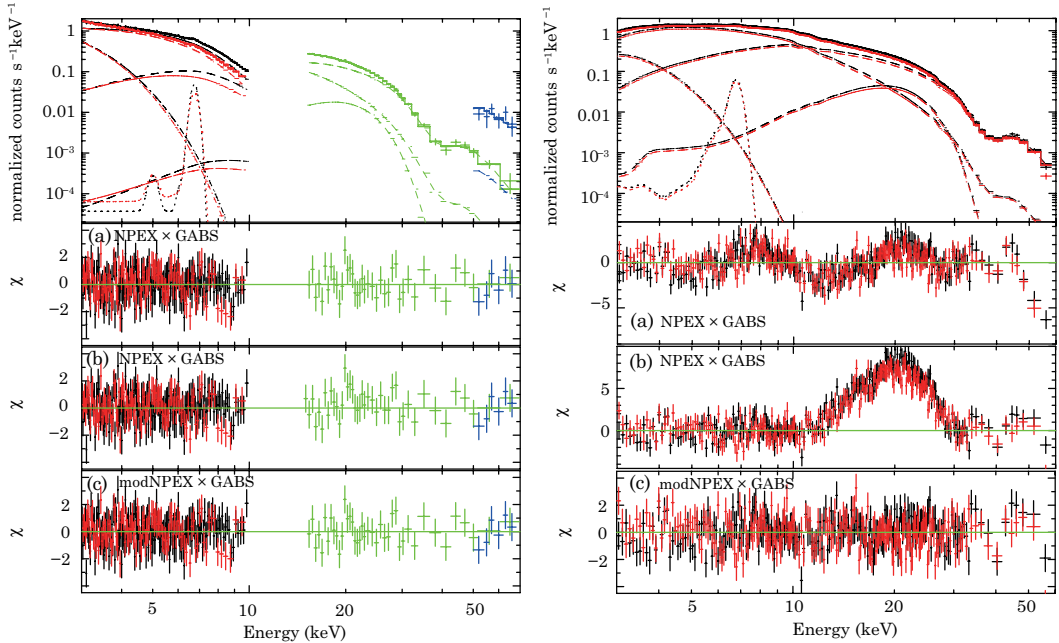}
\caption{Phase-averaged, background-subtracted spectra of
  4U\,1626$-$67 obtained with \textit{Suzaku} (left panel; black, red,
  green and blue crosses denote the XIS-FI, XIS-BI, HXD-PIN and
  HXD-GSO, respectively) and \textit{NuSTAR} (right panel; black and
  red crosses indicate the FPMA and FPMB, respectively). The top panel
  on either side shows the count rate spectra along with the best-fit
  models NPEX-with-broad-Gaussian-continuum model (modNPEX, dashed
  line) with a Gaussian absorption (GABS) model, Fe K$\alpha$ line
  (dotted line), and an additional black body (dashed-dotted line)
  model (see text and Table~\ref{table1}). The lower three panels show
  the residuals in units of $\chi$ from fitting the data with (a) NPEX with GABS, (b) Same
  model as (a) but the 10--30\,keV region was ignored when fitting the
  model, and (c) modNPEX with GABS.}
\label{fig5}
\end{figure*}

\subsection{Phase-averaged spectral analysis} \label{sec:avspec}

\subsubsection{Crab ratio}\label{sec:avgcrab}

First, we examine the ratio of the spectra to that of the Crab Nebula
and pulsar in order to get a model independent view of their overall
properties. Figure~\ref{fig4} shows the ratio of the phase-averaged
spectra obtained by \textit{Suzaku} and \textit{NuSTAR} to that of the
Crab, using a canonical model based on the \textit{Suzaku} calibration
results\footnote{Using the fits for XIS-nominal pointings without XIS2 found in Section 2.2.2 of \url{http://www.astro.isas.jaxa.jp/suzaku/doc/suzakumemo/suzakumemo-2008-06.pdf}}
as a template for the Crab spectrum. We fixed the
cross-normalization factor between XIS0 and the HXD-PIN to the standard value of
1.16. According to the \textit{NuSTAR} calibration results of \citet{mad17},
the cross-normalization factor between the FPMA and \textit{Suzaku}'s instruments are
$0.99\pm0.01$ for XIS0 and $1.13\pm0.03$ for the HXD-PIN, respectively. Comparing with these results,
the cross normalization uncertainty is at the few-percent level.

The CRSF is clearly visible at $\sim$40\,keV. The ratios in the soft
X-rays below 20\,keV show that the X-ray flux obtained with
\textit{NuSTAR} is brighter than that obtained with \textit{Suzaku},
which is consistent with the long-term X-ray variation observed by
MAXI (Figure~\ref{fig1}). Above 30\,keV, however, the hard X-ray flux
decreased in the latter observation. In addition, the spectral peak
energy appears to have slightly changed from $\sim$22.5\,keV in the
2010 \textit{Suzaku} observation to $\sim$21.5\,keV in the 2015
\textit{NuSTAR} observation. A similar spectral difference is also
reported by D17, who compared the long-term Swift/BAT spectrum with
the \textit{NuSTAR} data. D17 suggest that the difference is due to a
small change of the depth of the fundamental CRSF; however, the
model-independent Crab ratios in Figure~\ref{fig4} indicate that changes in the
continuum spectral shape should also be considered.

\begin{table}
\begin{center}
\caption{\label{table1}Best-fit parameters for the phase-averaged \textit{Suzaku} and \textit{NuSTAR} spectra of 4U 1626$-$67 using the modNPEX continuum models with one GABS component to describe the fundamental cyclotron line. }
\setlength{\tabcolsep}{2.0pt}
\begin{tabular}{lccc}
\hline
\hline
  & \multicolumn{2}{c}{modNPEX $\times$ GABS} \\

\hline
  &  \textit{Suzaku} & \textit{NuSTAR} \\
\hline
$kT_{\mathrm{BB}}$ (keV) &$ 0.47 ^ { + 0.06 } _ { -0.06 } $ &$ 0.47 ^ { + 0.04 } _ { -0.04 } $ \\
$A_{\mathrm{BB}}$\tablenotemark{a} &$ 258 ^ { + 271 } _ { -104 } $ &$ 254 ^ { + 174 } _ { -85 } $ \\
$\alpha$&$0.53^{+0.16}_{-0.18}$ &$0.47^{+0.09}_{-0.10}$ \\
$kT_{\mathrm{NPEX}} $ (keV) &$ 7.1 ^ { + 0.6 } _ { -0.4 } $ & $ 5.9 ^ { + 0.1 } _ { -0.1 } $  \\
$A_\mathrm{n}(\times 10^{-2})$\tablenotemark{b} &$2.3^{+0.7}_{-0.6}$ & $2.6^{+0.4}_{-0.4}$    \\
$A_\mathrm{p}(\times 10^{-5})$\tablenotemark{b} &$4.0^{+1.4}_{-1.8}$ &  $7.0^{+0.9}_{-1.3}$   \\
$E_{\mathrm{Gaussian}}$ (keV) &19.8(fix) & $19.8^{+0.4}_{-0.5}$  \\
$\sigma_{\mathrm{Gaussian}} $ (keV) &4.8(fix) &$4.8^{+0.6}_{-0.5}$\\
$A_{\mathrm{Gaussian}}(\times 10^{-3}) $\tablenotemark{d} &$1.8^{+2.3}_{-1.7}$ & $3.1^{+1.2}_{-0.8}$\\
$E_\mathrm{Fe}$ (keV)&$6.75^{+0.06}_{-0.06} $ & $6.76^{+0.05}_{-0.05} $ \\
$\sigma_\mathrm{Fe}$ (keV)&$0.13^{+0.06}_{-0.08}$ & $0.15^{+0.06}_{-0.06}$ \\
$A_\mathrm{Fe}(\times 10^{-4})$\tablenotemark{c} &$1.3^{+0.5}_{-0.4}$ &$1.2^{+0.3}_{-0.2}$\\
$E_\mathrm{CRSF}$ (keV) &$38.2^{+0.9}_{-0.8}$ &$37.7^{+0.1}_{-0.1}$   \\
$\sigma_\mathrm{CRSF} $ (keV) &$5.1^{+0.7}_{-0.7}$ &$4.2^{+0.1}_{-0.1}$ \\
$d_\mathrm{CRSF} $ &$22.9^{+5.7}_{-2.6}$ &$14.7^{+0.8}_{-0.8}$   \\
$C_\mathrm{XIS_{BI}}$\tablenotemark{d} &0.941$^{+0.007}_{-0.007}$&---&\\
$C_\mathrm{FPMB}$\tablenotemark{e} &---& 1.002$^{+0.002}_{-0.002}$\\
$\chi ^{2}$ / d.o.f&323.48/316 & 533.08/422	\\
$p_\mathrm{null}$\tablenotemark{f} & 3.7$\times$10$^{-1}$& 1.9$\times$10$^{-4}$\\
\hline
\end{tabular}
\end{center}
\tablenotetext{a}{$R_{\rm km}^{2}/d_{10}^{2}$, where $R_{\rm km}$ is the radius of the black-body in km and $d_{10}$ is the distance in units of 10\,kpc.}
\tablenotetext{b}{Normalization of the power-law. Defined at 1 keV in units of photons keV$^{-1}$ cm$^{-2}$ s$^{-1}$.}
\tablenotetext{c}{Normalization of the Gaussian. Defined in units of photons keV$^{-1}$ cm$^{-2}$ s$^{-1}$.}
\tablenotetext{d}{A cross calibration constant for XIS-BI relative to XIS-FI.}
\tablenotetext{e}{A cross calibration constant for FPMB relative to FPMA.}
\tablenotetext{f}{Null hypothesis probability.}
\end{table}

\subsubsection{Continuum}\label{sec:avcont}

Next, we modeled the phase-averaged spectra using the Negative and Positive power-law times EXponential (NPEX) model
\citep{miha95,maki99}. To model the CRSF, we multiplied the continuum by a Gaussian optical depth absorption model
(GABS in XSPEC) $\exp(-S(E))$, where
\begin{equation} \label{eq:gabs}
S(E) = \frac{d_\mathrm{a}}{\sqrt{2\pi}\sigma_\mathrm{a}} e^{-(E-E_\mathrm{a})^2 / (2\sigma_\mathrm{a}^2)},
\end{equation}
and where $E_\mathrm{a}$, $\sigma_\mathrm{a}$, and
$d_\mathrm{a}$ are the energy, width, and line depth. We also apply an iron line modeled by an additive Gaussian, and
an additional black body for the soft excess \citep{sch01} modeled by a
BBODYRAD model, to the \textit{Suzaku} and \textit{NuSTAR} data sets
separately. The hydrogen column density is fixed at the total Galactic H I
column density ($N_{\rm H} = 1 \times
10^{21}\,\mathrm{cm}^{-2}$)\footnote{\url{https://heasarc.gsfc.nasa.gov/cgi-bin/Tools/w3nh/w3nh.pl}}.

The NPEX model succeeded to reproduce the \textit{Suzaku} data (left of Figure \ref{fig5}a, with $\chi^2 /\mathrm{d.o.f.} = 326.08/317$) but failed to reproduce the \textit{NuSTAR} data (right of Figure \ref{fig5}a, with $\chi^2 /\mathrm{d.o.f.} = 1005.1/427$). 
However, we note that it can describe the
\textit{NuSTAR} spectrum quite well if the energies around the
spectral peak (between 10 and 30\,keV) are excluded.
Figure~\ref{fig5}b displays the results of a fit when the 10--30\,keV
data are ignored. This motivates the construction of a new model,
which we term ``modNPEX,'' consisting of an NPEX continuum with an
additional broad Gaussian emission feature around 20\,keV. Similar
bumps have been seen in other accretion-powered pulsars
\citep{cob02,klochkov07,ferrigno09,vasco13,farinelli16,Ballhausen17}. When fitting
the \textit{Suzaku} data, the center energy and the width of the
Gaussian model were fixed at the best-fit values of the
\textit{NuSTAR} data. We found that both data sets are well reproduced
by this model (Figure~\ref{fig5}c). A broad Fe K$\alpha$ emission line
is also detected at $\sim$6.76\,keV with a $\sigma$ of $\sim$150\,eV in both the
observations. These line centroid energies and widths are consistent with the
results in previous papers \citep[][\textit{Suzaku} data; D17, \textit{NuSTAR}
data]{cam12}. We also tried different phenomenological continuum models, the Fermi-Dirac CutOff (FDCO) model \citep{tanaka1986} and a power-law modified with a high-energy cut-off (HIGHECUT) model \citep{whi83,cob02,furst13}. As a result, we found that the modNPEX is the most successful in providing a good and consistent description of both the phase-averaged and the two phase-resolved spectra (Section~\ref{sec:phasespec}). Therefore, we adopted only the modNPEX as the continuum emission model in this paper.

It should be noted here that D17, using a HIGHECUT model \citep{whi83,cob02,furst13} or bulk+thermal
Comptonization continuum model \citep{becker07}, noted
absorption-like residuals around $\sim$60\,keV. They therefore added a second
Gaussian absorption model to describe this feature, which they interpret as the
harmonic of the CRSF. However, as D17 acknowledge, the centroid energy of the
feature deviates significantly from an integer multiple of the fundamental
energy, quite a bit more than one expects for a harmonic CRSF \citep[even if
relativistic corrections are taken into account, the harmonic should lie at
roughly twice the energy of the fundamental; see][]{katja05}. Additionally, the
line parameters are model-dependent: D17 find a significantly lower energy of
$61\pm1$\,keV when using the \citet{becker07} physical continuum model,
compared to their HIGHECUT fits ($67\pm3$\,keV), while the line depth drops from
$50^{+17}_{-10}$ to $22\pm5$.

In our analysis of the \textit{NuSTAR} spectra, a second GABS feature is not needed to obtain an acceptable fit.
Additionally, even when residuals around 60\,keV are present, they are limited
to a single bin at the extreme upper end of the useful \textit{NuSTAR}
spectrum. Thus, the feature reported by D17 is mainly constrained by the
\textit{Swift}-BAT spectrum. We further note that the \textit{Swift}-BAT has a
tagged $^{241}$Am calibration source on board, which emits 59\,keV photons
\citep{gehrels04}; this contributes to its background and could result in
line-like features. Thus, further study of 4U~1626$-$67 with energy coverage
considerably higher than 60\,keV is needed to properly study any possible
harmonic CRSF.

As the Crab ratios from Section~\ref{sec:avgcrab} suggest, our spectral
modeling shows a
decrease in the pseudo plasma temperature $kT$ from $7.1^{+0.6}_{-0.4}$\,keV to
$5.9\pm0.1$\,keV. While the CRSF is narrower and shallower in the \textit{NuSTAR}
observation compared to \textit{Suzaku}.
The unabsorbed X-ray fluxes in the 3.0--10.0\,keV, 30.0--60.0\,keV,
and 3.0--60.0\,keV bands derived with the modNPEX model fitting of the
\textit{Suzaku} data are $3.21^{+0.09}_{-0.12} \times 10^{-10}$,
$1.72^{+0.31}_{-0.28} \times 10^{-10}$, and
$1.29^{+0.14}_{-0.14} \times
10^{-9}\,\mathrm{erg}\,\mathrm{cm}^{-2}\,\mathrm{s}^{-1}$,
respectively. Assuming a distance of 10\,kpc \citep[optical observations
constrain its distance to 5--13\,kpc][]{cha98}, these fluxes correspond to an
X-ray luminosity of $1.54^{+0.17}_{-0.17}\times
10^{37}\,\mathrm{erg}\,\mathrm{s}^{-1}$ in the 3.0--60.0\,keV band. The
\textit{NuSTAR} fluxes in the same energy ranges are $3.44^{+0.04}_{-0.06}
\times 10^{-10}$, $1.14^{+0.07}_{-0.08} \times 10^{-10}$, and
$1.25^{+0.05}_{-0.04} \times
10^{-9}\,\mathrm{erg}\,\mathrm{cm}^{-2}\,\mathrm{s}^{-1}$, corresponding to an
X-ray luminosity of $1.50^{+0.06}_{-0.05}\times
10^{37}\,\mathrm{erg}\,\mathrm{s}^{-1}$ in the 3.0--60.0\,keV band. Comparison
the observed X-ray fluxes by \textit{Suzaku} with the \textit{NuSTAR} results,
we found that the 3.0--10.0\,keV flux increased by $\sim$5\%, whereas the
30.0--60.0\,keV flux decreased by $\sim$35\%.

\begin{figure}\centering
\includegraphics[width=7.5cm]{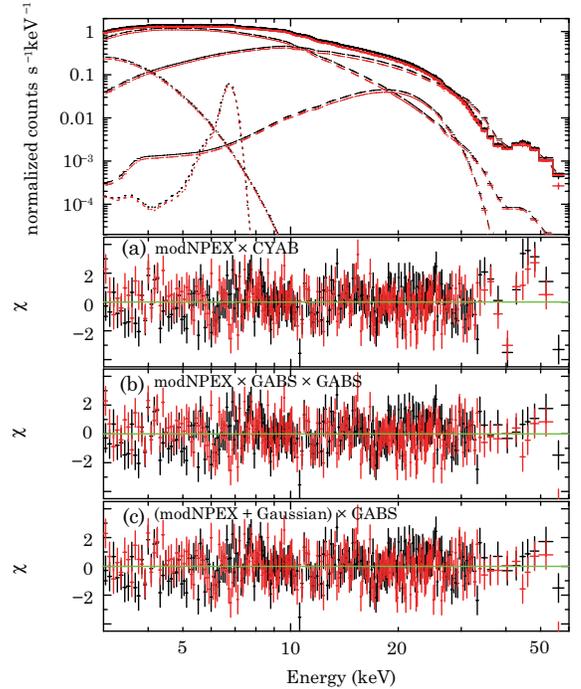}
\caption{Same \textit{NuSTAR} data as in Figure~\ref{fig5} right
  panels, but different models are applied. The top panel shows the
  phase-averaged spectrum with the best-fit model, consisting of a
  modNPEX continuum with two GABS features, both modeling the
  fundamental CRSF (see text and Table~\ref{table2}). The lower three
  panels show the residuals in units of $\chi$ from fitting with (a) modNPEX$\times$CYAB, (b) modNPEX with $2\times$GABS, and (c) modNPEX with
  GABS plus an additional Gaussian.}
\label{fig7}
\end{figure}

\begin{table*}
\begin{center}
\caption{\label{table2}Best-fit parameters for the phase-averaged \textit{NuSTAR} spectra of 4U 1626$-$67 using the modNPEX continuum models with different one- and two-component models to describe the fundamental cyclotron line.}
\setlength{\tabcolsep}{2.0pt}
\begin{tabular}{lcccc} 
\hline 
\hline 
  & modNPEX  & modNPEX  & (modNPEX + Gaussian) \\
  & $\times$ CYAB  & $\times$ GABS $\times$ GABS  & $\times$ GABS \\
\hline 
$kT_{\mathrm{BB}}$ (keV)  &$ 0.47 ^ { + 0.04 } _ { -0.04 } $  &$ 0.47 ^ { + 0.03 } _ { -0.03 } $   &$ 0.47 ^ { + 0.03 } _ { -0.03 } $\\
$A_{\mathrm{BB}}$\tablenotemark{a}  &$ 263 ^ { + 167 } _ { -89 } $  &$ 266 ^ { + 154 } _ { -86 } $  & $ 266 ^ { + 155 } _ { -87 } $\\
$\alpha$  &$0.52^{+0.06}_{-0.07}$  &$0.48^{+0.05}_{-0.05}$ &$0.48^{+0.05}_{-0.05}$\\
$kT_{\mathrm{NPEX}} $ (keV)   &$ 6.3 ^ { + 0.1 } _ { -0.1 } $   & $ 5.9 ^ { + 0.1 } _ { -0.1 } $   & $ 5.9 ^ { + 0.1 } _ { -0.1 } $\\
$A_n(\times 10^{-2})$\tablenotemark{b}  &$2.6^{+0.3}_{-0.4}$  & $2.6^{+0.2}_{-0.2}$    & $2.6^{+0.2}_{-0.2}$ \\
$A_p(\times 10^{-5})$\tablenotemark{b}  &$6.1^{+0.7}_{-1.0}$   &  $7.1^{+0.4}_{-0.4}$  &  $7.1^{+0.4}_{-0.4}$  \\
$E_{\mathrm{Gaussian}}$ (keV)  & $20.6^{+0.4}_{-0.6}$  & 19.8(fix)   & 19.8(fix)\\
$\sigma_{\mathrm{Gaussian}} $ (keV)  &$5.0^{+0.4}_{-0.4}$  &4.8(fix)  &4.8(fix)\\
$A_{\mathrm{Gaussian}}(\times 10^{-3}) $\tablenotemark{c}  &$3.1^{+1.3}_{-0.8}$ & 3.1$^{+0.3}_{-0.1}$  & 3.1$^{+0.3}_{-0.1}$\\
$E_\mathrm{Fe}$ (keV)  &$6.76^{+0.05}_{-0.05}$  & $6.76^{+0.05}_{-0.05} $  & $6.76^{+0.05}_{-0.05} $\\
$\sigma_\mathrm{Fe}$ (keV) &$0.15^{+0.06}_{-0.06}$  & $0.15^{+0.06}_{-0.06}$  & $0.15^{+0.06}_{-0.06}$\\
$A_\mathrm{Fe}(\times 10^{-4})$\tablenotemark{c} &$1.2^{+0.3}_{-0.2}$  &$1.2^{+0.3}_{-0.2}$  &$1.2^{+0.3}_{-0.2}$\\
$E_\mathrm{CRSF}$ (keV)  &$36.6^{+0.1}_{-0.1}$  &$39.9^{+0.6}_{-1.3}$    &$39.2	^{+0.8}_{-0.7}$\\
$\sigma_\mathrm{CRSF} $ (keV)   &$5.3^{+0.3}_{-0.3}$  &$2.6^{+0.6}_{-0.8}$  &$2.9^{+0.5}_{-0.5}$\\
$\tau_\mathrm{CRSF} $  &$1.6^{+0.1}_{-0.1}$  &---  &---  \\
$d_\mathrm{CRSF} $  &---  &$5.3^{+5.9}_{-3.8}$   &$8.8^{+2.8}_{-3.2}$ \\
$E_\mathrm{abs}$ (keV)  &---  &$35.9^{+1.4}_{-2.7}$  &$32.8^{+0.9}_{-1.1}$  \\
$\sigma_\mathrm{abs} $ (keV)   &---  &$3.6^{+0.5}_{-0.7}$  &$3.4^{+0.5}_{-0.5}$\\
$d_\mathrm{abs} $  &---  &$8.7^{+4.2}_{-5.6}$  &---  \\
$A_{\mathrm{abs}}(\times 10^{-3}) $\tablenotemark{c}  &---  &---  &$-1.2^{+0.4}_{-0.4}$\\
$C_\mathrm{FPMB}$\tablenotemark{d}  & 1.002$^{+0.002}_{-0.002}$  & 1.002$^{+0.002}_{-0.002}$  & 1.002$^{+0.002}_{-0.002}$\\
$\chi ^{2}$ / d.o.f  &596.91/422  & 508.08/421  & 507.98/421\\
$p_\mathrm{null}$\tablenotemark{e}  & 2.1$\times$10$^{-9}$  & 2.3$\times$10$^{-3}$   & 2.3$\times$10$^{-3}$\\
\hline 
\end{tabular} 
\end{center} 

\tablenotetext{a}{$R_{\rm km}^{2}/d_{10}^{2}$, where $R_{\rm km}$ is the radius of the black-body in km and $d_{10}$ is the distance in units of 10\,kpc.}
\tablenotetext{b}{Normalization of the power-law. Defined at 1 keV in units of photons keV$^{-1}$ cm$^{-2}$ s$^{-1}$.}
\tablenotetext{c}{Normalization of the Gaussian. Defined in units of photons keV$^{-1}$ cm$^{-2}$ s$^{-1}$.}
\tablenotetext{d}{A cross calibration constant for FPMB relative to FPMA.}
\tablenotetext{e}{Null hypothesis probability.}
\end{table*}

\subsubsection{An asymmetric cyclotron line}\label{sec:avgcrsf}

The residuals of the modNPEX fit to the \textit{NuSTAR}
data show a slight systematic structure remaining at around the
fundamental CRSF energy (Figure~\ref{fig5}c, right).
The presence of a complex line profile for the fundamental CRSF is
consistent with D17, despite their different approach to modeling the spectrum.
We tried different models to reproduce this feature better and investigate its
shape. First, we replaced the GABS model by a pseudo-Lorentzian absorption
model (CYAB model, \texttt{cyclabs} in XSPEC) and fitted the spectrum using
modNPEX as a continuum model. As can be seen in Figure~\ref{fig7}
and Table~\ref{table2}, this model did not improve the fit relative to the GABS
CRSF model. Next, we added another GABS model, as has been used for V\,0332+53
(X\,0331+53) \citep{krey05,katja05,nakajima10} and Cep X-4 \citep{furst15}.
With this model, we successfully eliminated the residuals around the CRSF
(Figure~\ref{fig7} and Table~\ref{table2}). These results are largely
consistent with D17, although they tie the energies of the GABS models to be
the same.

However, since the GABS model only produces an absorption feature, we also
tried replacing the second GABS component with an additive Gaussian model
(i.e., $[\mbox{modNPEX}+\mbox{Gaussian}]\times\mbox{GABS}$), allowing the Gaussian
normalization to be both positive and negative. The best-fit normalization
using the model is negative at the $5\sigma$ level, and the energy was
consistent with that found using the $2\times\mbox{GABS}$ CRSF. To evaluate the
chance probability of improvement of adding the extra Gaussian component, we
simulated 400,000 data sets using \texttt{simftest} in XSPEC. The evaluated
chance probability was $6.3\times 10^{-5}$. Therefore, we have possibly detected a
complex line profile for the fundamental CRSF in the phase-averaged spectrum.

\begin{figure*}
  \includegraphics[width=\textwidth]{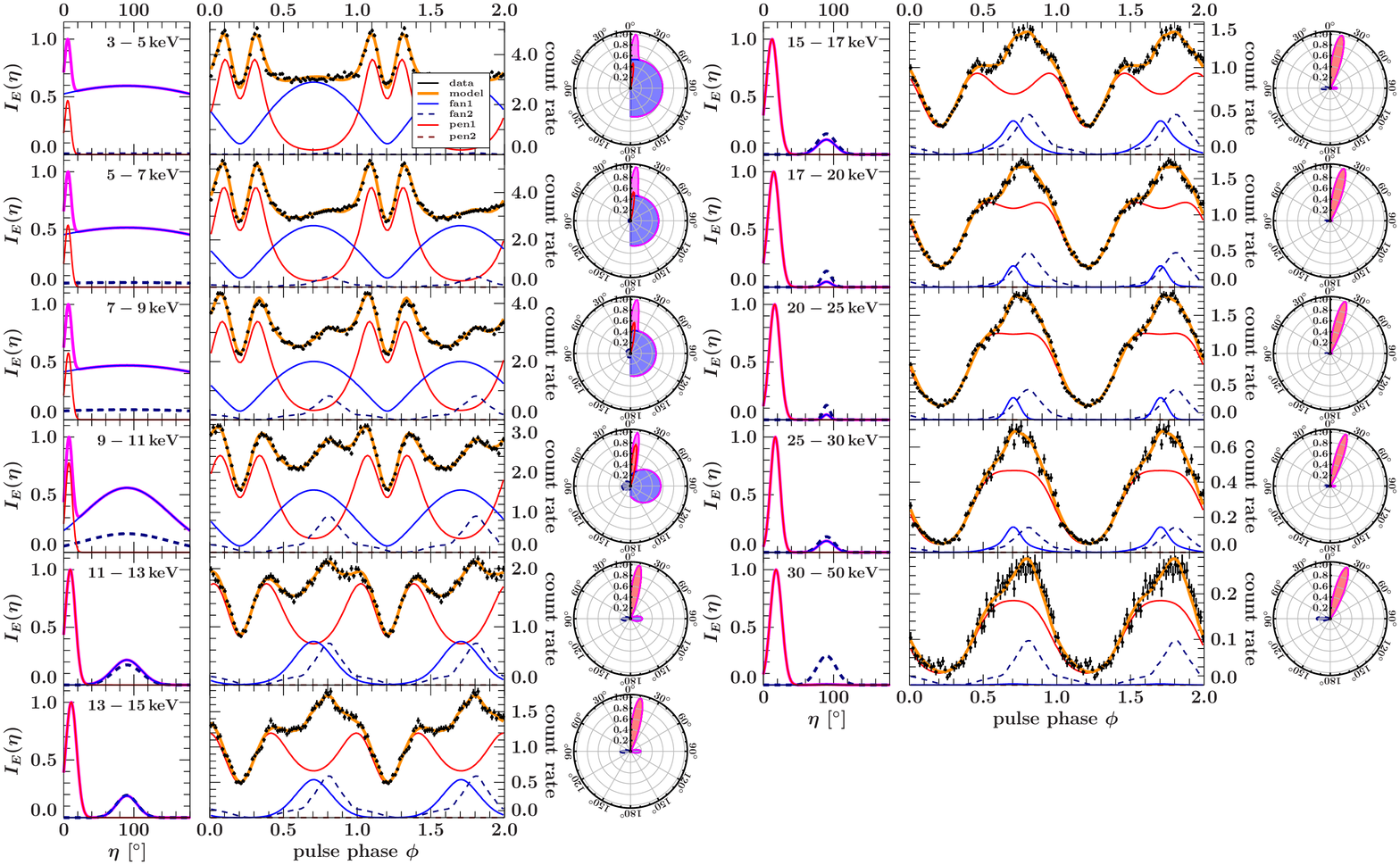}
  \caption{Modeled emission patterns and pulse profiles of {4U\,1626$-$67} in
    different energy bands. The center panels show the pulse profiles obtained
    with \textit{NuSTAR} (black points) and the fitted model (orange) with its
    individual components of the fan (solid blue, dashed navy) and pencil
    emission (solid red, dashed maroon) of the first and second accretion
    column, respectively. The left-hand panels show the corresponding
    normalized emission patterns, $I_E(\eta)$, of the two accretion columns.
    The solid magenta line corresponds to the combined emission pattern of the
    fan and pencil beam of the first column. The right-hand panels show the
    same emission patterns as polar plot, where the right and left side
    counting $\eta$ clockwise and counter-clockwise correspond to the first and
    second accretion column, respectively. The best-fit parameters are shown in
    Table~\ref{tab:pptab}.}
  \label{fig8}
\end{figure*}

\begin{figure}
  \includegraphics[width=\columnwidth]{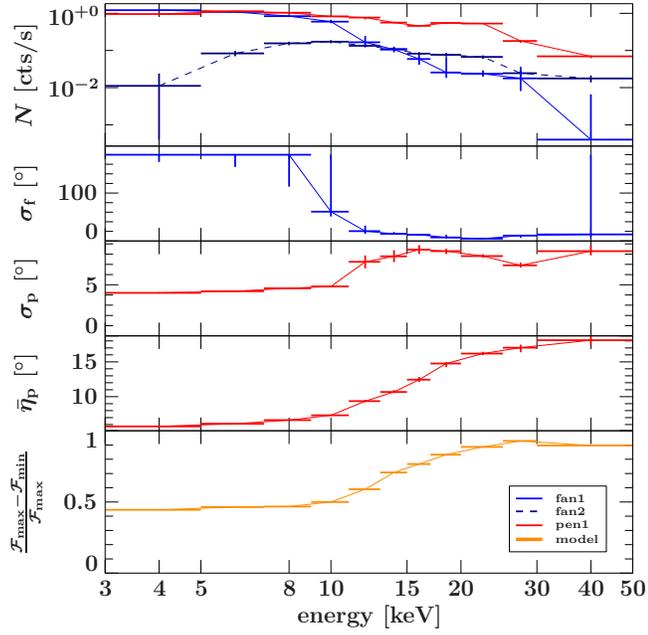}
  \caption{Energy dependent fit parameters of the accretion column model, with 
the model's pulsed fraction in the bottom panel. See Table~\ref{tab:pptab} for 
the best-fit parameters. The panels showing $N$, $\sigma_\mathsf{f}$, 
$\sigma_\mathsf{p}$, and $\bar{\eta}_\mathsf{p}$ include statistical 
errors (Table~\ref{tab:pptab}). Note that the emission angle for the fan beam is 
fixed at $\bar{\eta}_\mathrm{f}=90^\circ$.}
  \label{fig:ppevol}
\end{figure}

\begin{figure}
  \includegraphics[width=\columnwidth]{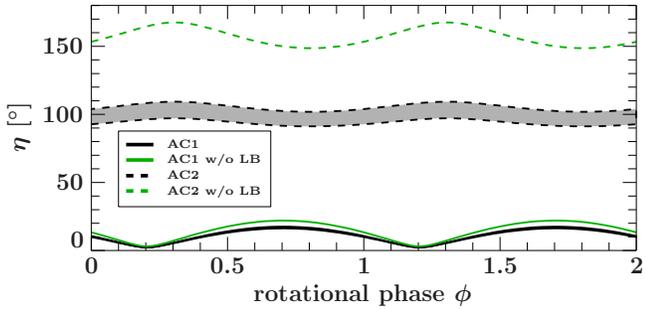}
  \caption{Phase dependent angle, $\eta$, under which the magnetic
    field is seen at either pole according to the best-fit values in
    Table~\ref{tab:pptab}. Solid and dashed black lines enclose $\eta$
    values corresponding to the first and second accretion column, AC1
    and AC2, respectively, accounting for light bending. Green lines
    show the case of neglecting light bending.}
  \label{fig:pptheta}
\end{figure}

\begin{table*}
\caption{Fit parameters of the accretion column model. Parameters not listed 
here are fixed or tied. In particular we impose
$r_\mathrm{AC1}=r_\mathrm{AC2}$ and
$h_\mathrm{AC1}=h_\mathrm{AC2}$ for the column size,
$N_{\mathrm{p}_2}=N_{\mathrm{p}_1}$,
$\sigma_{\mathrm{p}_2}=\sigma_{\mathrm{p}_1}$,
$\bar{\eta}_{\mathrm{p}_2}=\bar{\eta}_{\mathrm{p}_1}$ for the pencil beam,
and $\sigma_{\mathrm{f}_2}=\sigma_{\mathrm{f}_1}$ and
$\bar{\eta}_{\mathrm{f}_2}=\bar{\eta}_{\mathrm{f}_1} = 90^\circ$ for the
fan beam. The given errors correspond to the 90\% confidence level.}
\label{tab:pptab}
\centering
\begin{tabular}{ c | rrrrr | rr | r}
 global & \multicolumn{5}{|c|}{column$_1$} & \multicolumn{2}{|c|}{column$_2$} & 
{} \\
 $i$    & $\Phi_\mathrm{AC1}$ & $\Theta_\mathrm{AC1}$ & $r_\mathrm{AC1}$ & 
$h_\mathrm{AC1}$ & {} & $\Phi_\mathrm{AC2}$ & $\Theta_\mathrm{AC2}$ & 
$\chi^2$/d.o.f. \\
{}[deg] & [deg] & [deg] & [m] & [m] & {} & [deg] & [deg] &  {}\\
\hline
$9.5\pm0.2$ & $73.6\pm0.3$ & $12.50^{+0.07}_{-0.11}$ 
& $590\pm50$ & $252^{+12}_{-14}$ & {} & 
$290\pm2$ & $158.1\pm0.2$ & $1146/631$\\
\hline\hline
{}  & \multicolumn{3}{|c}{pencil$_1$} & \multicolumn{2}{c|}{fan$_1$} & 
\multicolumn{2}{|c|}{fan$_2$} &  {}\\
 $\Delta E$ & $N_{\mathrm{p}_1}$ & $\sigma_{\mathrm{p}_1}$ & 
$\bar{\eta}_{\mathrm{p}_1}$ & $N_{\mathrm{f}_1}$     & $\sigma_{\mathrm{f}_1}$ & 
\multicolumn{2}{|c|}{$N_{\mathrm{f}_2}$} &  \\
 {[keV]}    & [cts/s] & [deg]      & [deg] & [cts/s] & [deg] &  
\multicolumn{2}{|c|}{[cts/s]} &  {}\\
\hline
$3$ -- $5$ & $0.95^{+0.01}_{-0.02}$ & $4.21^{+0.07}_{-0.06}$ & $5.73\pm0.06$ 
& $1.21^{+0.03}_{-0.02}$ & $180^{+0}_{-16}$ & 
\multicolumn{2}{|c|}{$0.011^{+0.013}_{-0.011}$} & \\
$5$ -- $7$ & $1.14\pm0.02$ & $4.37\pm0.06$ & $6.12\pm0.06$ & 
$1.09^{+0.03}_{-0.02}$ & $180^{+0}_{-26}$ & 
\multicolumn{2}{|c|}{$0.08\pm0.02$} & \\
$7$ -- $9$ & $1.03^{+0.03}_{-0.02}$ & $4.66\pm0.05$ & $6.65^{+0.07}_{-0.06}$ 
& $0.84^{+0.03}_{-0.04}$ & $180^{+0}_{-67}$ & 
\multicolumn{2}{|c|}{$0.16\pm0.02$} & \\
$9$ -- $11$ & $0.83^{+0.06}_{-0.03}$ & $4.84^{+0.05}_{-0.06}$ & 
$7.33^{+0.10}_{-0.08}$ & $0.60^{+0.08}_{-0.06}$ & $61^{+120}_{-10}$ & 
\multicolumn{2}{|c|}{$0.17\pm0.02$} & \\
$11$ -- $13$ & $0.77^{+0.04}_{-0.09}$ & $7.3\pm0.7$ & $9.4\pm0.2$ & 
$0.17^{+0.08}_{-0.04}$ & $20^{+12}_{-6}$ & 
\multicolumn{2}{|c|}{$0.13\pm0.02$} & \\
$13$ -- $15$ & $0.56^{+0.02}_{-0.03}$ & $7.8\pm0.6$ & $10.7^{+0.3}_{-0.2}$ 
& $0.11^{+0.03}_{-0.02}$ & $14.8^{+4.0}_{-0.7}$ & 
\multicolumn{2}{|c|}{$0.11\pm0.02$} & \\
$15$ -- $17$ & $0.46^{+0.03}_{-0.02}$ & $8.4\pm0.5$ & $12.4\pm0.4$ & 
$0.06\pm0.02$ & $13.0^{+0.9}_{-1.7}$ & 
\multicolumn{2}{|c|}{$0.082\pm0.009$} & \\
$17$ -- $20$ & $0.55^{+0.02}_{-0.04}$ & $8.3\pm0.3$ & 
$14.8^{+0.3}_{-0.6}$ & $0.026^{+0.050}_{-0.008}$ & $7^{+6}_{-3}$ & 
\multicolumn{2}{|c|}{$0.08\pm0.01$} & \\
$20$ -- $25$ & $0.53\pm0.01$ & $7.8\pm0.2$ & 
$16.2\pm0.3$ & $0.024^{+0.006}_{-0.005}$ & 
$5.0^{+1.7}_{-0.3}$ & \multicolumn{2}{|c|}{$0.067^{+0.007}_{-0.008}$} & \\
$25$ -- $30$ & $0.18^{+0.02}_{-0.03}$ & $6.9\pm0.3$ & 
$17.0^{+0.5}_{-0.7}$ & $0.02^{+0.02}_{-0.01}$ & $11^{+2}_{-5}$ & 
\multicolumn{2}{|c|}{$0.025\pm0.005$} & \\
$30$ -- $50$ & $0.070^{+0.004}_{-0.008}$ & $8.3^{+0.6}_{-0.5}$ & $18.1\pm0.7$ & 
$\le0.007$ & $14^{+167}_{-8}$ & \multicolumn{2}{|c|}{$0.018\pm0.004$} &\\
\end{tabular}

\end{table*}

\subsection{Pulse profile modeling}\label{sec:pulseprofile}

Because the energy spectrum of 4U\,1626$-$67 is known to depend strongly on
pulse phase \citep{pra79,kii86,iwa12}, we present in these next two sections a
detailed phase-resolved study. First, to investigate the geometry of the
neutron star quantitatively, we performed pulse profile modeling using a new
relativistic ray tracing code. Due to the low time resolution of the
\textit{Suzaku} data below 10\,keV, we concentrate on the \textit{NuSTAR} data.
Figure~\ref{fig8} shows the energy-resolved and background subtracted pulse
profiles in eleven different energy bands obtained with \textit{NuSTAR}. The
pulse profiles strongly depend on energy, with a double-peaked structure below
10\,keV which becomes single-peaked and almost sinusoidal in the higher energy
bands. The characteristics of the pulse profiles seen by \textit{NuSTAR} are
consistent with the \textit{RXTE} results observed in 2010 \citep{beri14}.

We use the relativistic light bending code of \citet[][see also
\citealt{schoenherr2014}]{falkner2017a,falkner2017b} to model the energy-resolved
pulse profiles of 4U\,1626$-$67 (see Figure~\ref{fig8}). This code
follows a similar approach to \cite{ferrigno2011} to obtain the
observable energy and phase dependent flux. In contrast to
\cite{ferrigno2011}, we are able to apply arbitrary emission patterns
to emission regions of any geometrical shape. The code has been used
previously by \citet{schoenherr2014} to investigate the energy
dependent phase lags in accreting neutron stars.

In our model for 4U\,1626$-$67 we consider a canonical neutron star of
mass $M=1.4M_\odot$ and radius $R = 10$\,km. The observed X-rays are
emitted by two cylindrical accretion columns AC1 and AC2 of height
$h_\mathrm{AC1,AC2}$ and radius $r_\mathrm{AC1,AC2}$. Allowing for an
asymmetric magnetic field, the columns are positioned individually at
azimuthal angles $\Phi_\mathrm{AC1,AC2}$ and polar angles
$\Theta_\mathrm{AC1, AC2}$, respectively, in a coordinate system that
is measured with respect to the neutron star's rotational axis. The
angle between the line of sight and the neutron star's angular
momentum vector specifies the inclination $i$ of the neutron
star. Hence $i=0^\circ$ would correspond to a face-on system.

We make the simplified assumption that the emission pattern of the
columns can be described as a mixture of Gaussian-like fan and pencil
beam emission components in the frame of rest of the neutron star's
surface, e.g., at a given energy the emissivity of one accretion
column is given by
\begin{equation} \label{eq:ppint}
 I_E(\eta) = N_\mathrm{p}
\exp{\left(-\left[\frac{\eta-\bar{\eta}_\mathrm{p}}{\sqrt{2}\sigma_\mathrm{p
} } \right]^2\right)} +
N_\mathrm{f}
\exp{\left(-\left[\frac{\eta-\bar{\eta}_\mathrm{f}}{\sqrt{2}\sigma_\mathrm{f
} }\right]^2\right)}\quad,
\end{equation}
where $\eta$ is the angle of the emitted photons measured with respect to the 
magnetic field axis in the frame of rest of the neutron star surface, and where 
the energy dependent quantities $\bar{\eta}$, $\sigma$, and $N$ describe the 
direction of peak emissivity, the width, and the strength of the pencil-beam 
($p$) and fan-beam ($f$) components, respectively. For the fan beam we set 
$\bar{\eta}_\mathrm{f}=90^\circ$, i.e., the fan beam is fixed to emit from the 
sides of the accretion column perpendicular to the B-Field. While we impose 
the same model for the emissivity pattern on both poles, the fluxes of the beams 
are allowed to vary freely.

We model the energy-dependence of the pulse profile by letting the parameters 
$N$, $\sigma$, and $\bar{\eta}$ of the emission pattern be energy dependent. For 
the models described in the following we assume that the emissivity of the 
accretion column is independent of height and thus constant over the whole 
column. From Equation~\ref{eq:ppint} we then derive the observed energy- and 
phase- dependent total flux 
\begin{equation}\label{eq:ppflux} 
\mathcal{F}_E(\phi) = F_1(\phi, I_E) + F_2(\phi, I_E)\quad,
\end{equation}
where $\phi$ is the pulse phase and $F_{1,2}$ is the flux of the individual 
accretion columns emitting with the given, energy dependent, emission pattern 
$I_E$ (Eq.~\ref{eq:ppint}). The description of the calculation of $F_{1,2}$, 
which accounts for all general relativistic effects and is in addition to the 
parameters here dependent on the neutron star's inclination, $i$, is beyond the 
scope of this paper and is given in \cite{falkner2017a,falkner2017b}.

Figure~\ref{fig8} includes the best-fit model described above for the 
energy-resolved and background subtracted \textit{NuSTAR} pulse profiles in 11 
energy bands, which were fitted simultaneously. Lines show the overall model 
and the individual contributions of the fan and pencil beams from each 
accretion column. The corresponding parameters are listed in 
Table~\ref{tab:pptab}. The geometrical parameters ($i$, 
$\Theta_\mathrm{AC1,AC2}$, $\Phi_\mathrm{AC1,AC2}$, $h_\mathrm{AC1,AC2}$, 
$r_\mathrm{AC1,AC2}$) are global, that is the same for all energies, while the 
parameters describing the emission profile are determined for each individual 
energy band. In the best-fit case the size of the second accretion column is 
tied to the first one, i.e, $r_\mathrm{AC2}=r_\mathrm{AC1}$ and 
$h_\mathrm{AC2}=h_\mathrm{AC1}$. Further, the shape of the energy dependent 
emission profiles of the two columns are tied together, i.e., 
$\sigma_{\mathrm{p}_2}=\sigma_{\mathrm{p}_1}$, 
$\bar{\eta}_{\mathrm{p}_2}=\bar{\eta}_{\mathrm{p}_1}$ for the pencil beam 
component, and $\sigma_{\mathrm{f}_2}=\sigma_{\mathrm{f}_1}$ and 
$\bar{\eta}_{\mathrm{f}_2}=\bar{\eta}_{\mathrm{f}_1} = 90^\circ$ for the fan 
beam component. We found that there are two local minima of the $\chi^2$ landscape of the observer inclination $i$; one solution is $i = 9^\circ$, the other is $i = 27^\circ$. Since the $9^\circ$ solution shows a simpler pulse profile evolution than the $i = 27^\circ$ solution and because of its consistency with the physical simulation results taking into account the anisotropy of the scattring cross section in a strong magnetic field according to \citet{kii86} (see details in Section \ref{sec:discuss_pulseprofile}), we only show the results from the $i = 9^\circ$ solution in this paper. Table~\ref{tab:pptab} also gives the 90\% confidence levels for 
the parameters. These uncertainties are purely statistical and driven by a 
complex $\chi^2$ landscape. It is probable that these uncertainties are 
systematically underestimated. In relation to each other the uncertainties, 
however, indicate that the radius and height of the column, and for some 
energies the width of the fan-beam are much less constraint than the other 
parameters. 

The derived best-fit values of $\Theta_\mathrm{AC1,AC2}$, 
$\Phi_\mathrm{AC1,AC2}$ indicate an asymmetric B-Field configuration of 4U 
1626$-$67. Such an orientation of a magnetic field axis was also suggested in 
the previous pulse profile modeling results of \citet{leahy91} which shows that 
nine of 20 pulsars requires an magnetic field axis offset. The fit shows that 
the magnetic field of the first column passes through close to the line of sight 
during each rotation. That is, at pulse phase $\phi=\Phi_\mathrm{AC1}$, when the 
first column is in the front, we are looking at the first column from above with 
an angle to its magnetic field axis of approximately 
$\Theta_\mathrm{AC1}-i=3\degr$. Furthermore, the displacement of the two 
accretion columns compared to the symmetric antipodal case is given  
\begin{equation}\label{eq:displacement}
\Delta = \pi - \arccos\left( \vec{n}_\mathrm{AC1}\circ\vec{n}_\mathrm{AC2} 
\right)\quad,
\end{equation}
which is the angular distance of the two unit vectors, 
$\vec{n}_{\mathrm{AC1},\mathrm{AC2}}$, of the accretion columns positions 
corresponding to their azimuthal and polar angles. The geometry of our best-fit 
model yields a displacement of $\Delta=13\fdg9$, which represents a moderate 
asymmetry. In our best-fit model we tie the accretion columns dimensions to 
prevent parameter degeneracy. Such an asymmetric B-field configuration, however, 
may suggest columns of different sizes due to possible asymmentric accretion 
flows, that is different accretion rates for the two poles \citep[see 
e.g.,][]{becker07,postnov15}. Therefore, and due to the fact that we make the 
very simple assumption of an homogeneously emitting surface, the obtained 
heights and radii of the columns should not be interpreted as physical 
quantities.

Despite the simplified assumptions entering the beam pattern, our model 
describes the observed pulse profiles and their energy evolution remarkably well 
and with a smooth variation of all relevant parameters of the emission 
characteristics. The evolution of the pulse profile is characterized by a very 
wide fan beam which strongly decreases in width as the energy increases. In 
contrast, the shape of the pencil beam component only slightly changes, but the 
direction of its peak emission does. This shift explains the widening of the gap 
between the double peak in the pulse profile. From 3\,keV to 10\,keV the 
parameters corresponding to the first column change only slightly, but the 
normalization of the fan beam of the second column increases explaining the 
changes seen in the pulse profile at these lower energies. For the geometry in 
our best-fit model the pencil beam component of the second accretion column is 
directed away from the observer over the whole rotational phase and is therefore 
never observed. To ensure the correctness\footnote{Otherwise the fit-algorithm 
might be stuck at a $N_{\mathrm{p}_2}=0$ solution, disregarding the second 
pencil beam solution also for other tested geometries.} of our fit we therefore 
tie the normalization of the two pencil beam components, i.e., 
$N_{\mathrm{p}_2}=N_{\mathrm{p}_1}$. Figure~\ref{fig:ppevol} shows the energy 
dependency of the parameter values.

The decomposition of the pulse profiles in the middle columns of Figure~\ref{fig8} 
illustrates how these parameter changes manage to reproduce the pulse profiles 
so well: The pencil beam is responsible for the distinct and symmetric double 
peak that characterizes the softer energy bands. The peaks are close together 
since the pencil beam is directed upwards, with only a small offset, 
$\bar{\eta}_\mathrm{p}$, to the magnetic field, which resembles a conical 
emission pattern. The strong non-pulsed continuum between the two peaks is 
produced by the broad fan beam, which is shifted by half a phase with respect to 
the dip between the double peak of the pencil beam. With increasing energy the 
double peak decreases in importance and its width broadens while the gap in 
between increases. This behavior is reflected in the best-fit parameters by 
showing that the offset angle $\bar{\eta}_\mathrm{p}$ and beam width 
$\sigma_\mathrm{p}$ increase with energy (see Figure~\ref{fig:ppevol}). In 
addition, the flat plateau at low energies evolves into an asymmetric peak, 
which is caused by the narrowing of the fan beam. The asymmetry visible in this 
pulse is caused by the fan beam of the slightly-misaligned second accretion 
column. A consequence of this best-fit geometry is that the pencil beam of the 
second column is directed away from the observer at all pulse phases, and thus 
it is not observable. A second consequence is that the evolution of the 
direction of the peak emissivity of the primary pencil beam, 
$\bar{\eta}_\mathrm{p_1}$, dominates the evolution of the pulsed fraction, 
$(\mathcal{F}_\mathrm{max}-\mathcal{F}_\mathrm{min})/\mathcal{F}_\mathrm{max}$, 
while the width of the fan beam, $\sigma_\mathrm{f}$, has only a minor influence 
(Fig.~\ref{fig:ppevol}).

We note that extrapolating the behavior of the pencil and fan beams to
even lower energies than considered here predicts an evolution of the
pulse profile towards a shape dominated by the single broad hump of
the fan beam, consistent with the pulse profiles seen by
\textit{XMM-Newton} \citep{kra07} and \textit{Chandra}
 \citep{hemphill17}.

Modeling the pulse profiles also yields our viewing angle, $\eta$,
onto the two accretion columns. This parameter is important for the
interpretation of the CRSF, the shape of which strongly depends on the
angle under which we see the magnetic field \citep{sch17a,sch17b}. As
shown in Figure~\ref{fig:pptheta}, $\eta$ is strongly influenced by
the effect of light bending, for the first column $\eta$ varies
between $5^\circ$ and $22^\circ$ in a small band with mean width $\sim
1^\circ$, whereas for the second column the mean width of the band is
$\sim 11^\circ$ between $91^\circ$ and $110^\circ$.

Compared to other models put forward for explaining the energy
dependent change of the pulse profile of 4U\,1626$-$67, our pulse
decomposition explains the observed energy dependent behavior solely
by a change in the emission characteristics of the accretion column,
without invoking foreground effects, such as the absorption by an accretion stream proposed by \citet{beri14}. The
simpler explanation is possible by virtue of the low inclination of
$i=9\fdg5$, where relativistic effects allow a complex interplay
between the pencil and fan beam to produce the observed profiles.
Although there is some systematic uncertainty in the derived inclination angle due to the complexity of the pulse profile modeling, we note that the
inclination is in reasonable agreement with the face on inclination of
$\lesssim8^\circ$ inferred by studies of the orbit of the system that
assume that the donor star is a $0.08\,M_\sun$ hydrogen-depleted and
partially degenerate star \citep{levine88,verbunt90a,cha98}.

The inclination is in moderate disagreement, however, with the $i
\lesssim 33^{\circ}$ estimate for a $0.02\,M_\sun$ helium or
carbon-oxygen white dwarf donor \citep{verbunt90a,cha98}. This
higher-inclination case is supported by the presence of a complex of
broad, double-peaked emission lines around 1\,keV \citep{sch01,kra07},
which are consistent with an inclination in the range of
30--40$^{\circ}$ \citep{schulz13,hemphill17}. The low inclination
found by our pulse profile modeling can possibly be reconciled with
the high inclination implied by the disk lines if the angular momenta
of the accretion disk and neutron star are misaligned. This would
result in a strong warp in the accretion disk, which could explain the
disk flips that have been invoked to explain the torque reversals of
4U~1626$-$67 \citep{kerkwijk98,wijers99}.

\begin{figure}\centering
\includegraphics[width=6.0cm, angle=270]{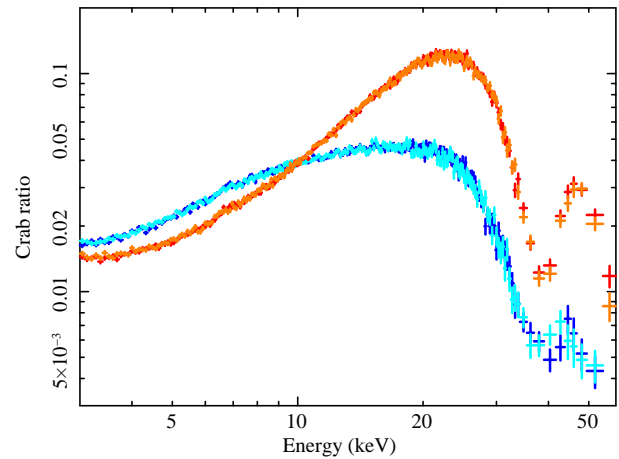}
\caption{The phase-resolved energy spectra normalized with a canonical
  model of the Crab (Crab Nebula and pulsar) spectrum. Red and orange
  crosses show the FPMA and FPMB spectra extracted from phase interval
  $\phi=0.5$--$1.0$ in Figure~\ref{fig8}, while blue and cyan crosses
  show those from phase interval $\phi=0.0$--$0.5$.}
\label{fig11}
\end{figure}
\subsection{Phase-resolved spectral analysis}\label{sec:phasespec}

Due to the strong angular dependency of the cyclotron scattering cross
section, we expect that the continuum emission and CRSF profile will
depend on our viewing direction \citep[see, e.g.,][]{mes92}. Here we
perform a phase-resolved analysis to investigate how the spectral
parameters change with phase. Phase-resolved spectroscopy of the
\textit{NuSTAR} data was explored in D17, however, their 20 phase bins had poor
photon statistics, which precluded a detailed study of the fundamental CRSF.
Motivated by our pulse profile modeling results, we instead divide the events
into two phase intervals: $\phi = $0.0--0.5, which is more dominated by soft
X-ray flux and contains the double peaks seen at low energies, and 0.5--1.0,
where the profile is flat at low energies and more dominated by hard X-ray
flux. Based on our pulse profile modeling, the phase 0.0--0.5 interval
corresponds to viewing angles $\eta$ of the first accretion column between
$5^{\circ}$ and $16^{\circ}$, while the 0.5--1.0 interval corresponds to angles
between $17^{\circ}$ and $22^{\circ}$.

\subsubsection{Crab ratio}\label{sec:phasecrab}

Similar to our treatment of the phase averaged data, we first performed a model
independent study of the phase dependent data using Crab ratios. The normalized
phase-resolved spectra by the canonical model of the Crab are shown in
Figure~\ref{fig11}. It is clear that the continuum emission and CRSF profile
both depend strongly on the spin phase, with phases 0.5--1.0 being considerably
harder spectrally than the 0.0--0.5 interval.

\begin{figure}
\includegraphics[width=7.0cm]{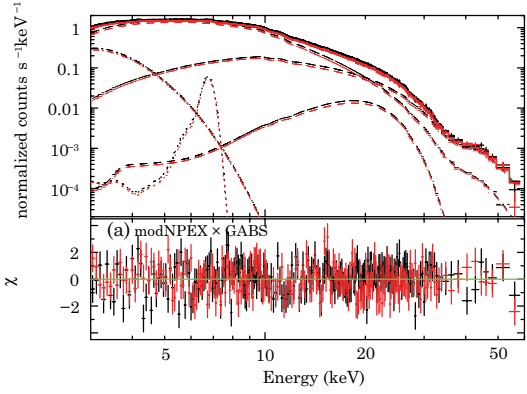}
\caption{Phase-resolved spectra extracted from the phase interval
  $\phi=0.0$--$0.5$ as defined in Figure~\ref{fig8} (black and red
  crosses for FPMA and FPMB, respectively) with the best-fit model
  consisting of modNPEX with a GABS component. The lower panels show
  $\chi$ residuals of the best-fit results for the models of (a)
  modNPEX with GABS.}
\label{fig12}
\end{figure}

\begin{figure}
\includegraphics[width=7.0cm]{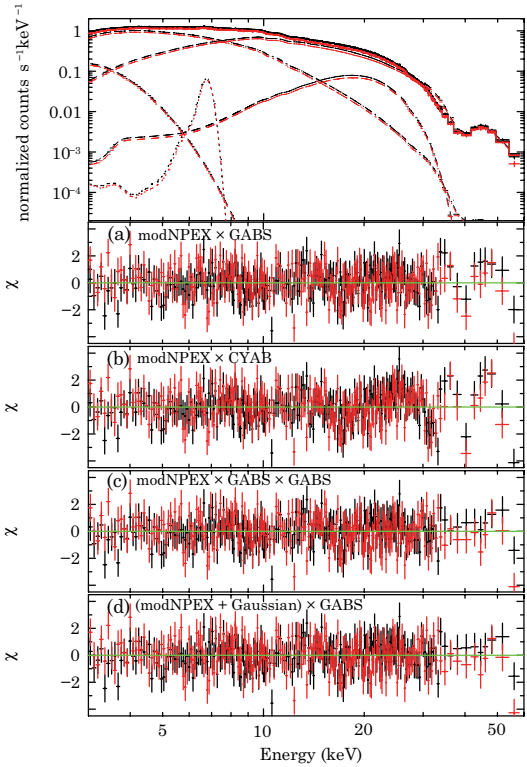}
\caption{Phase-resolved spectra extracted from the phase interval
  $\phi=0.5$--$1.0$ as in Figure~\ref{fig8} (black and red crosses
  for FPMA and FPMB, respectively) with the best-fit model consisting
  of modNPEX with two GABS components. The lower panels show $\chi$
  residuals of the best-fit results using (a) modNPEX with GABS, (b) modNPEX with CYAB, (c) modNPEX with
  two GABS, and (d) modNPEX with GABS plus an additional Gaussian.}
\label{fig13}
\end{figure}

\begin{figure}
\includegraphics[width=7.0cm]{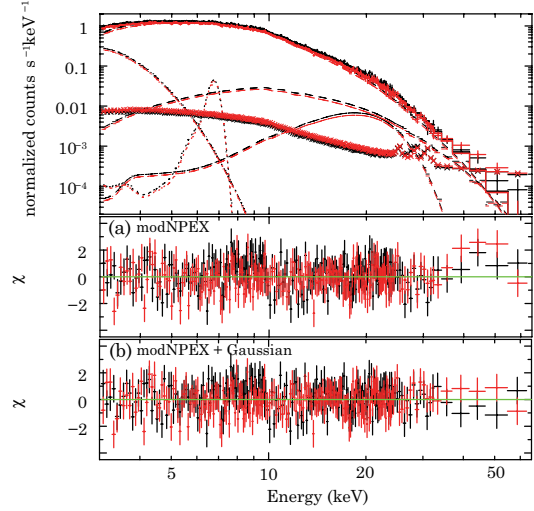}
\caption{Phase-resolved spectra extracted from the phase interval
  $\phi=0.125-0.250$ defined in Figure~\ref{fig8} (black and red
  crosses for FPMA and FPMB, respectively) with the best-fit modNPEX
  model. The background spectra are shown as well (black and red
  x-symbols for FPMA and FPMB, respectively). The lower panels show
  the $\chi$ residuals for the best-fit models with (a) a pure modNPEX
  continuum and (b) modNPEX with an additional Gaussian.}
\label{fig14}
\end{figure}

\subsubsection{Continuum and cyclotron line}\label{sec:phasespec_fit}

We model the phase-resolved spectra with the modNPEX
continuum models, which is the successful in reproducing the
phase-averaged spectra (Section~\ref{sec:avcont}). Due to the lower
signal to noise ratio of the spectra, the iron line energy and width and those of broad Gaussian component for the modNPEX
model were fixed at the values from the phase-averaged spectrum
(Table~\ref{table1}).

Figure~\ref{fig12} shows the results of the
spectral fitting of the spectrum accumulated for phase interval
$\phi=0.0$--$0.5$. We do
obtain a good fit with the modNPEX continuum, and find that a single GABS
component (Figure~\ref{fig12}) is sufficient to reproduce the CRSF profile. 

We then fitted the spectrum from phase interval $\phi=0.5$--$1.0$ with
the same model. The results are displayed in Figure~\ref{fig13}.
With a single GABS component, V-shaped residuals are still
visible around the fundamental CRSF (see Figure~\ref{fig13}a), similar
to our results for the phase-averaged \textit{NuSTAR} spectrum in
Section~\ref{sec:avgcrsf}. As before, we tried a CYAB model (a
pseudo-Lorentzian profile) in place of the GABS component, but this produced
worse fits compared to the GABS CRSF, with~$\chi^2 /\mathrm{d.o.f.} =
651.69/427$ (Figure~\ref{fig13}b). We then tried another GABS component (i.e.,
modNPEX$\times$GABS$\times$GABS) and an additive Gaussian component
([modNPEX$+$Gaussian]$\times$GABS). In both cases, the extra residuals around
the CRSF were eliminated (Figure~\ref{fig13}c,d), and the Gaussian component
normalization is negative at the $5\sigma$ level. The probability of this
feature arising by chance, determined by simulating 400,000 datasets with
\texttt{simftest} in XSPEC, is $1.8\times10^{-5}$. The best fit parameters using the modNPEX model are summarized in Table~\ref{table4}.

Hence, we conclude that we have again tentatively detected the
distorted CRSF which cannot be represented by a simple Gaussian or
pseudo-Lorentzian absorption model, similar to the CRSF seen in the
phase-averaged \textit{NuSTAR} spectrum.

\begin{table*} 
\begin{center}
\caption{\label{table4}Best-fit parameters for the phase 0.0--0.5 and 0.5--1.0 \textit{NuSTAR} spectra
 of 4U\,1626$-$67.}
\setlength{\tabcolsep}{2.0pt}
\begin{tabular}{l|c|cccccc}
\hline
\hline
 & modNPEX & modNPEX & modNPEX & (modNPEX + Gaussian) \\
 & $\times$ GABS & $\times$ GABS & $\times$ GABS $\times$ GABS & $\times$ GABS\\
\hline
Pulse phase $\phi$&0.0--0.5& \multicolumn{3}{c}{0.5--1.0}\\
\hline
$kT_{\mathrm{BB}}$ (keV) & $ 0.49 ^ { + 0.04 } _ { -0.04 } $ &$ 0.42 ^ { + 0.07 } _ { -0.07 } $ &$ 0.42 ^ { + 0.07 } _ { -0.07 } $ & $ 0.42 ^ { + 0.07 } _ { -0.07 } $\\
$A_{\mathrm{BB}}$\tablenotemark{a} &$ 244 ^ { + 158} _ { -85 } $ &$ 352 ^ { + 1165 } _ { -214 } $ &$ 359^ { + 1026 } _ { -210 } $ &$ 367 ^ { + 1079 } _ { -217 } $ \\
$\alpha$ & $0.47^{+0.01}_{-0.01}$ &$1.02^{+0.08}_{-0.09}$ &$1.00^{+0.08}_{-0.10}$ &$1.02^{+0.08}_{-0.09}$\\
$kT_{\mathrm{NPEX}} $ (keV)  & $ 5.6 ^ { + 0.3 } _ { -0.2 } $&$ 6.0 ^ { + 0.1 } _ { -0.1 } $  &$ 6.0 ^ { + 0.1 } _ { -0.1 } $ &$ 6.0 ^ { + 0.1 } _ { -0.1 } $ \\
$A_n(\times 10^{-2})$\tablenotemark{b} &$ 1.9 ^ { + 0.2 } _ { -0.2 } $ &$4.8^{+0.6}_{-0.7}$ &$4.7^{+0.7}_{-0.7}$ &$4.9^{+0.7}_{-0.7}$   \\
$A_p(\times 10^{-5})$\tablenotemark{b} &$ 3.2 ^ { + 1.1 } _ { -1.2 } $ &$10.3^{+0.3}_{-0.3}$  &$12.6^{+0.3}_{-0.3}$  & $10.4^{+0.3}_{-0.3}$ \\
$E_{\mathrm{Gaussian}}$ (keV) & 19.8(fix) & 19.8(fix)  & 19.8(fix) & 19.8(fix) \\
$\sigma_{\mathrm{Gaussian}} $ (keV) &4.8(fix) & 4.8(fix) & 4.8(fix) & 4.8(fix)\\
$A_{\mathrm{Gaussian}}(\times 10^{-3}) $ &$1.1^{+0.3}_{-0.3}$  &$5.1^{+0.3}_{-0.3}$ &$5.5^{+0.3}_{-0.3}$ &$5.4^{+0.3}_{-0.3}$\\
$E_{Fe}$ (keV) &6.76(fix) & 6.76(fix) & 6.76(fix) &6.76(fix)\\
$\sigma_{Fe}$ (keV)& 0.15(fix) &0.15(fix) &0.15(fix) &0.15(fix)\\
$A_\mathrm{Fe}(\times 10^{-4})$\tablenotemark{c} &$1.2^{+0.3}_{-0.3}$ &$1.3^{+0.3}_{-0.3}$ &$1.3^{+0.3}_{-0.3}$&$1.3^{+0.3}_{-0.3}$\\
$E_a$ (keV) &$36.5^{+0.5}_{-0.4}$ &$37.9^{+0.1}_{-0.1}$ &$39.2^{+1.0}_{-0.7}$ &$38.9^{+0.5}_{-0.5}$  \\
$\sigma_\mathrm{CRSF} $ (keV)  &$4.6^{+0.4}_{-0.4}$ &$4.0^{+0.1}_{-0.1}$ &$2.8^{+0.5}_{-0.5}$ &$3.0^{+0.3}_{-0.3}$\\
$d_\mathrm{CRSF}$ &$10.8^{+1.7}_{-1.6}$ &$15.9^{+0.7}_{-0.7}$ &$10.3^{+2.8}_{-5.5}$ & $11.7^{+1.7}_{-2.1}$\\
$E_\mathrm{abs}$ (keV) &--- &--- &$34.2^{+2.2}_{-1.5}$ &$32.5^{+0.7}_{-0.8}$   \\
$\sigma_\mathrm{abs} $ (keV)  &--- &--- &$2.9^{+0.8}_{-0.5}$ &$2.8^{+0.5}_{-0.5}$ \\
$d_\mathrm{abs} $ &---&--- &$4.8^{+5.6}_{-2.6}$ &---  \\
Norm$_{abs} \times 10^{-3}$ &---&--- &--- &$-1.4^{+0.5}_{-0.5}$   \\

$\chi ^{2}$ / d.o.f &453.52/427 &522.50/427  &493.80/424 &494.55.47/424	\\
$p_\mathrm{null}$\tablenotemark{d} & 1.8$\times$10$^{-1}$& 1.1$\times$10$^{-3}$ & 1.1$\times$10$^{-2}$ & 1.0$\times$10$^{-2}$\\

\hline
\end{tabular}
\end{center}
\tablenotetext{a}{$R_{\rm km}^{2}/d_{10}^{2}$, where $R_{\rm km}$ is the radius of the black-body in km and $d_{10}$ is the distance in units of 10\,kpc.}
\tablenotetext{b}{Normalization of negative ($n$) and positive ($p$) NPEX components, defined at 1 keV in units of photons keV$^{-1}$ cm$^{-2}$ s$^{-1}$.}
\tablenotetext{c}{Normalization of the Gaussian. Defined in units of photons keV$^{-1}$ cm$^{-2}$ s$^{-1}$.}
\tablenotetext{d}{Null hypothesis probability.}
\end{table*}


\begin{table}
\begin{center}
\caption{\label{table5}Best fit parameters for the phase 0.125--0.250 \textit{NuSTAR} spectra
 of 4U 1626-67.}
\setlength{\tabcolsep}{2.0pt}
\begin{tabular}{lccccccc}
\hline
\hline
 & modNPEX & modNPEX \\
 & & + Gaussian emission \\
\hline
$kT_{\mathrm{BB}}$ (keV) & $ 0.41 ^ { + 0.05 } _ { -0.05 } $ &$ 0.47 ^ { + 0.05 } _ { -0.05 } $ \\
$A_{\mathrm{BB}}$\tablenotemark{a} &$ 684 ^ { + 1250} _ { -397 } $ &$ 324 ^ { + 424 } _ { -159 } $   \\
$\alpha$ & $0.10^{+0.10}_{-0.09}$ &$-0.15^{+0.18}_{-0.04}$ \\
$kT_{\mathrm{NPEX}} $ (keV)  &$ 5.3 ^ { + 0.5 } _ { -0.5 } $ &$ 4.6 ^ { + 6.0 } _ { -0.6 } $    \\
$A_n(\times 10^{-2})$\tablenotemark{b} &$1.6^{+0.2}_{-0.2}$  &$1.3^{+0.3}_{-0.2}$     \\
$A_p(\times 10^{-5})$\tablenotemark{b} &$0.5^{+1.1}_{-0.4}$ &$<4.9$     \\
$E_{\mathrm{Gaussian}}$ (keV) &　19.8(fix)  & 19.8(fix)    \\
$\sigma_{\mathrm{Gaussian}} $ (keV) &4.8(fix) & 4.8(fix)\\
$A_{\mathrm{Gaussian}}(\times 10^{-3}) $ &$0.5^{+0.2}_{-0.2}$  &$1.0^{+0.3}_{-0.3}$\\
$E_{Fe}$ (keV) &6.76(fix) & 6.76(fix)\\
$\sigma_{Fe}$ (keV)& 0.15(fix) &0.15(fix) \\
$A_\mathrm{Fe}(\times 10^{-4})$\tablenotemark{c} &$0.8^{+0.5}_{-0.5}$ &$0.9^{+0.5}_{-0.5}$\\
$E_a$ (keV) &--- &40.5(fix)    \\
$\sigma $ (keV)  &--- &$8.5^{+3.8}_{-2.5}$  \\
$\mathrm{Norm} (\times 10^{-4})$&--- &$2.2^{+0.7}_{-0.7}$    \\	
$\chi ^{2}$ / d.o.f &395.68/384 &366.99/382   \\
$p_\mathrm{null}$\tablenotemark{d} & 3.3$\times$10$^{-1}$& 7.0$\times$10$^{-1}$ \\

\hline
\end{tabular}
\end{center}
\tablenotetext{a}{$R_{\rm km}^{2}/d_{10}^{2}$, where $R_{\rm km}$ is the radius of the black-body in km and $d_{10}$ is the distance in units of 10\,kpc.}
\tablenotetext{b}{Normalization of negative ($n$) and positive ($p$) NPEX components, defined at 1 keV in units of photons keV$^{-1}$ cm$^{-2}$ s$^{-1}$.}
\tablenotetext{c}{Normalization of the Gaussian. Defined in units of photons keV$^{-1}$ cm$^{-2}$ s$^{-1}$.}
\tablenotetext{d}{Null hypothesis probability.}
\end{table}

\subsubsection{Dim phase spectrum}\label{sec:phasespec_dim}

\citet{iwa12} reported the possible detection of an emission-line-like feature
in the dim phase ($\phi=0.125$--$0.250$ in Figure~\ref{fig8}) of 4U\,1626$-$67
during the spin-down state. We extracted a spectrum from this phase interval
and fitted it with the modNPEX model; the results are displayed in
Figure~\ref{fig14}a and Table~\ref{table5}. The spectrum is well reproduced by
only the continuum model --- the CRSF feature does not appear in the dim phase
significantly. However, the residuals around 45 keV do indicate a possible
emission-line-like feature, although the signal is weak compared with the
background level. If we add a Gaussian emission component in the same manner as
\citet{iwa12}, with the center energy fixed at their best-fit value
$E=40.5$\,keV (Figure~\ref{fig14}b), the fitted width of
$\sigma=8.5^{+3.8}_{-2.5}$\,keV is consistent with the width found in the
spin-down state, which showed $\sigma>4.5$\,keV, whereas the normalization of
$2.2^{+0.8}_{-0.7} \times 10^{-4}$ photons cm$^{-2}$ s$^{-1}$ is lower than the
spin-down observation's $8.8^{+7.2}_{-2.9}\times
10^{-4}$\,ph\,cm$^{-2}$\,s$^{-1}$. The chance probability of improvements of adding this emission feature is $2.0\times10^{-2}$ determined by simulating 20,000 datasets with
\texttt{simftest} in XSPEC. Therefore, we concluded that we have only marginally detected the emission line at about 2$\sigma$ level. The observed unabsorbed fluxes for the dim phase of the spin-down state and the spin-up state in the 3 - 60 keV band are $2.2^{+0.2}_{-0.2} \times 10^{-10}$ and 
$6.0^{+0.1}_{-1.0} \times 10^{-10}$ \,$\mathrm{erg}\,\mathrm{cm}^{-2}\,\mathrm{s}^{-1}$, respectively (When we calculate the fluxes, the blackbody parameters are fixed). We note that while D17 claimed the presence of an
absorption feature in the dim phase of the pulse, their chosen phase interval
is different from ours, corresponding to approximately $\phi=$0.2--0.35\ in
Figure~\ref{fig8}, and thus this discrepancy is not problematic.

\section{Discussion} \label{sec:discussion}
We have presented an analysis of two observations of 4U\,1626$-$67 with
\textit{Suzaku} in 2010 September and \textit{NuSTAR} in 2015 May. We have
performed broad-band spectral analysis, using both phase-averaged and
phase-resolved data. In the phase-averaged analysis, we have found a change in
the continuum shapes between the observations and confirmed the complex
profile of the fundamental CRSF in the \textit{NuSTAR} data suggested by D17.
In the phase-resolved analysis using the \textit{NuSTAR} data, we have shown
the phase dependence of the continuum and CRSF feature, and find further
evidence for a distorted CRSF profile during the brighter phase. We have also
modeled the energy-resolved pulse profiles using a new relativistic ray tracing
code. In this section, we discuss the nature of 4U\,1626$-$67, based on our
findings.

\subsection{Difference of the continuum and pulse period between the two
observations}\label{sec:discuss_spectral_var}

We have found that the continuum emission is different between the 2010
\textit{Suzaku} and 2015 \textit{NuSTAR} observations. Let us consider the
implications of this result.

According to \citet{GL79}, for accretion via a disk, the rate of change in the
pulse period, $\dot{P}$, is proportional to $\dot{M}^{6/7}$, where $\dot{M}$ is
the total mass accretion rate onto the neutron star. Thus, we can estimate the
accretion rate onto 4U\,1626$-$67 using $\dot{P}$ from \textit{Fermi}-GBM
monitoring. As found in Section~\ref{sec:variab}, $\dot{P}$ over the
\textit{Suzaku} observation was $-2.8\times10^{-11}$\,s\,s$^{-1}$, while over
the \textit{NuSTAR} observation it was $-3.3\times10^{-11}$\,s\,s$^{-1}$. The
15\% higher $\dot{P}$ during the \textit{NuSTAR} observation would thus imply a
13\% higher accretion rate per \citet{GL79}.

Based on the Crab ratios of these data sets (Figure~\ref{fig4}), as well as our
spectral fitting results (Table~\ref{table1}), hard X-ray photons are
suppressed in the \textit{NuSTAR} observation relative to the 2010
\textit{Suzaku} spectrum. Since hard X-ray photons are mainly produced by
thermal Comptonization in the accretion column \citep{becker07}, we can infer
that the electron temperature of the plasma decreased. However, the photon
index $\alpha$ did not change significantly between the observations.

Summarizing our interpretations of our timing and spectral results, the
\textit{NuSTAR} observation saw a higher accretion rate and lower temperature
in the accretion column compared to the \textit{Suzaku} observation. At least
in terms of these results, the spectral variation between the 2010
\textit{Suzaku} and 2015 \textit{NuSTAR} observations indicate that the
decrease in the plasma temperature is caused by the increased accretion rate.
However, the relation between the accretion rate and the temperature is not
simple due to the complex of radiative transfer processes under the strong
magnetic field. Future theoretical study is needed to verify our
interpretation.

\subsection{Origin of the continuum emission}\label{sec:discuss_continuum_emission}

The high sensitivity of \textit{NuSTAR} in the hard X-ray band and our analyses
of the phase-averaged and phase-resolved spectra lead us to conclude that the
most appropriate empirical model for the continuum emission of 4U\,1626$-$67 is
the modNPEX model. The results indicate that the extra Gaussian is needed to allow the NPEX model-based fit to provide a good description of the physical continuum. Similar broad Gaussian features have been detected from several other accretion-powered pulsars, e.g., 4U 0115+63, Cen X-3, Her X-1 and A 0535+26 \citep{ferrigno09,suchy08,vasco13,Ballhausen17}. Since the broad Gaussian feature appears around the spectral peak, it is inferred that the main difference between the simple NPEX continuum model and the observed spectrum is the spectral shape of the quasi-exponential cutoff. The NPEX model approximates the Wien peak like cutoff which is expected for pure thermal Comptonization regime but the shape of the spectral cutoff is mainly determined by the contribution ratio of the bulk and thermal Comptonization in their photon propagation process. Moreover, since there are three types of seed photons (bremsstrahlung and cyclotron emission created along the column, black-body emission from the base of the column), the cutoff shape is also related to the contribution ratio of the seed photons. Therefore, a physical model fitting is important to investigate the origin of the broad Gaussian feature observed from 4U\,1626$-$67. However, this is beyond the scope of the present paper. While D17 did fit 4U 1626-67 with the \citet{becker07} bulk and thermal Comptonization model, their description includes an additional reflection component and the unusual 2nd harmonic absorption feature described earlier. Because of this and since the individual spectral contributions of the different types of Comptonized seed photons are not shown, we cannot interpret the 20 keV residual within the D17 picture.

\begin{figure}\centering
\includegraphics[width=5.0cm, angle=270]{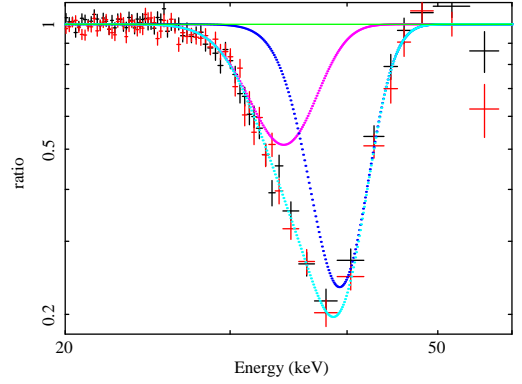}
\caption{Ratio residuals of the phase $\phi=0.5$--$1.0$ spectra to the
continuum component of the best-fit modNPEX model, with the two GABS components
excluded (FPMA:black crosses, FPMB:red crosses). Magenta, blue, and cyan dotted
lines show the contributions of the primary GABS component, secondary GABS
component, and sum of the two GABS models, respectively.}
\label{fig15}
\end{figure}

\begin{figure}\centering
\includegraphics[width=5.0cm, angle=270]{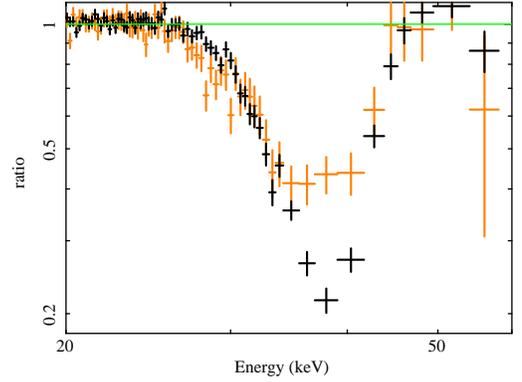}
\caption{Comparing CRSF shapes between phase $\phi$ = 0.5--1.0 and $\phi$ = 0.0--0.5. These are ratio residuals displayed in the same manner as in Figure~\ref{fig15}. Black crosses show the ratio residuals for the phase $\phi$ = 0.5--1.0 FPMA spectrum, while orange crosses show the residuals for phase $\phi$ = 0.0--0.5.}
\label{fig16}
\end{figure}

\begin{figure}\centering
\includegraphics[width=5.0cm, angle=270]{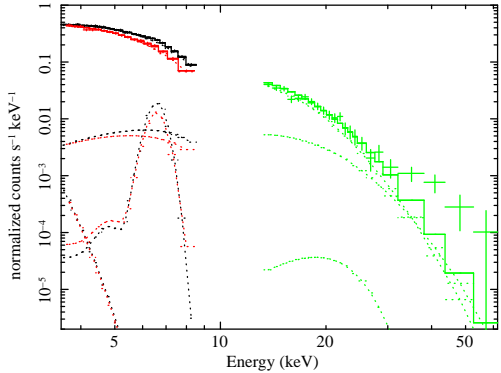}
\caption{The dim phase spectrum during the spin-down phase of 4U 
 1626$-$67 observed by \textit{Suzaku} in 2006 September 
 \citep{iwa12} with the modNPEX-based best-fit model. Black, red, and 
 green crosses are the XIS-FI, XIS-BI, and HXD-PIN data, 
 respectively. Histograms show the overall modNPEX-based best-fit 
 model and dotted lines show individual model components.}
\label{fig17}
\end{figure}

\subsection{Origin of the pulse profile evolution}\label{sec:discuss_pulseprofile}
Using our new relativistic ray tracing code, we successfully reproduce the
energy-resolved pulse profiles obtained by \textit{NuSTAR}
(Section~\ref{sec:pulseprofile}). We find that the beam pattern is
energy-dependent. The qualitative explanation for the observed pulse profile
evolution is found in the anisotropy of the Thomson scattering cross section in a strong magnetized plasma. In this case the cross
section depends on the photon energy and is different for the ordinary and
extraordinary polarization modes. \citet{kii86} simulated the 4U 1626$-$67
pulse profile during spin-up, as obtained by \textit{Tenma} during May 3--5,
1983. Their simulation is based on anisotropic radiation transfer calculations
by \citet{nag81}, ignoring vacuum polarization. For a cylindrical
geometry with a diameter of the Thomson scattering optical depth $\tau_{\mathrm{Th}} = 80$, \citet{kii86} find the
following: first, ordinary-mode photons dominate in lower energy band, whereas
extraordinary-mode photons become dominant toward higher energies. Second, the
maximum emission angle for extraordinary-mode photons is offset with respect to
the magnetic field, and that offset becomes slightly stronger at higher
energies. Third, the emission profile with respect to the magnetic field
becomes wider at higher energies. Comparing these with our pulse profile
modeling results, we can thus draw qualitative connections between our
fan-beam evolution and \citet{kii86}'s ordinary-mode emission profile, and
between our pencil-beam evolution and their extraordinary-mode emission
profile. Therefore, we suggest that the origin of the observed pulse profile
evolution is found in the energy and polarization dependence of the scattering
cross section in a strongly magnetized plasma. 

X-ray polarimetry will be useful to validate this hypothesis, as
ordinary- and extraordinary-mode photons contribute to polarization with
opposite signs.  Consider, as an example, the future mission
\textit{IXPE} \citep{Weisskopf16}, which will operate in the 2--8\,keV
band. Under our suggestion that the fan and pencil beams respectively
correspond to ordinary and extraordinary-mode photons, Figure~\ref{fig8}
predicts that in \textit{IXPE}'s band, the polarization degree will be
nearly zero around phase $\phi=0.45$, where the pencil and fan
contributions are comparable, and maximal around phases $\phi=0.3$ and
$\phi=0.75$, where one beam dominates over the other.  Meanwhile, the
hard X-ray band covered by, e.g., \textit{X-Calibur} \citet{Beilicke14}
should find higher overall polarization with a minimum around phase
$\phi=0.7$.

\subsection{CRSF profile}\label{sec:discuss_crsfprofile}

We have found that the observed fundamental CRSF in the phase-averaged and the
phase $\phi$ = 0.5--1.0 spectra is better described with a two-Gaussian
absorption structure than with a single Gaussian or pseudo-Lorentzian profile\
(Section~\ref{sec:phasespec_fit}).
To highlight this asymmetry, we show in Figure~\ref{fig15}
the ratio between the data and the continuum component of the best-fit model,
where we have excluded the CRSF model. The best-fit
two-Gaussian absorption model is overlaid, showing broadening at lower energies. This is similar to what was reported in
the \textit{NuSTAR} observation of Cep X-4 \citep{furst15}. Some explanations
of this distorted profile have been proposed. For example, \citet{nishimura11}
proposed that the fundamental line profile becomes asymmetric and shallower
toward lower energies due to the superposition of multiple CRSFs produced at
different altitudes along the column, assuming some gradient in the density,
temperature, and magnetic field. Another interpretation is photon spawning due
to inelastic scattering at higher harmonics \citep{sch07} --- as electrons
excited into higher Landau levels cascade down to the ground state, many
emitted photons will have similar energy to the fundamental, ``filling in'' the
fundamental line. In addition, our pulse profile modeling suggests that in the
phase interval $\phi$ = 0.5--1.0, the second accretion column contributes about
10\% of the flux of the first column. Thus, the emission from the second pole
may slightly contribute to the CRSF shape. However, a detailed study of the
spectra resulting from the mixing of light from the two columns is beyond the
scope of this work.

Our phase-resolved spectral analysis also suggests that the CRSF profile
depends on the spin phase. We have found that the CRSF in the phase $\phi$ =
0.0--0.5 spectra, which corresponds to the angle $\eta_\mathrm{AC1}$ =
5$^{\circ}$ -- 16$^{\circ}$ (Figure~\ref{fig:pptheta}), is well-reproduced by a
simple GABS model. Figure~\ref{fig16} shows a comparison of the CRSF shape
between phase $\phi$ = 0.5--1.0 and $\phi$ = 0.0--0.5 in terms of data-to-model ratio (with the absorption components
excluded). The depth of the line at $\phi$ = 0.0--0.5 is clearly shallower than at $\phi$ = 0.5--1.0. It is noteworthy that the wings of the two profiles resemble each
 other while the core of the $\phi$ = 0.5--1.0 profile appears to be
 filled in compared to the $\phi$ = 0.0--0.5 profile. To evaluate the width of the CRSF profiles, we calculate the FWHM
(full-width-at-half-maximum) from these ratios. The results are 12.9 keV and
11.4 keV corresponding to $\phi$ = 0.0--0.5 and $\phi$ = 0.5--1.0
, respectively. Therefore,
the CRSF is wider and shallower between $\eta_\mathrm{AC1}$ = 5$^{\circ}$ --
16$^{\circ}$ compared to $\eta_\mathrm{AC1}$ = 17$^{\circ}$ -- 21$^{\circ}$ (see Figure\ref{fig:pptheta}).
The relation between the CRSF shape and the angle $\eta$ is qualitatively
consistent with theoretical simulations when assuming a slab-type geometry
\citep{ise98,sch17b}. The relation is also qualitatively consistent with the
theoretical results assuming a cylindrical geometry illuminated by anisotropic
injections and magnetic field gradients \citep{nishimura15}.

The absorption feature disappears in the dim phase spectrum (phase
$\phi$=0.125--0.250). Although an emission-line-like feature　was detected in the \textit{Suzaku} spectrum observed during the 2006 spin-down state of
4U\,1626$-$67 \citep{iwa12}, it is only marginally detected in the \textit{NuSTAR} observation (the probability of the feature arising
by chance is 2.0 $\times 10^{-2}$). The fact indicates that the flux of the emission component decreased to the point of non-detectability in the 2015 spin-up state. According to our spectral fits (see
Section \ref{sec:phasespec_dim}), the intensity of the emission
feature during the \textit{NuSTAR} observation is about 4 times lower
than during the 2006 \textit{Suzaku} observation. On the other hand,
the 3--60\,keV flux of the \textit{NuSTAR} observation is about 3
times higher than that of the 2006 \textit{Suzaku} observation. As a consistency
check we refitted the 2006 \textit{Suzaku} observation with the same
modNPEX model used for the \textit{NuSTAR} observation in Section
\ref{sec:phasespec_dim} (Figure \ref{fig17}). We find consistent
results with the emission-line-like feature around 40\,keV being
possibly detected (the probability of this feature arising by chance is $8.0 \times 10^{-4}$ using 20,000 datasets with \texttt{simftest}) and the fitted continuum approximating the
exponentially cut-off power-law used by \citet{iwa12} since the
contribution of the broad Gaussian and the positive exponential
component are comparatively small
($A_{\mathrm{Gaussian}}<0.4\times10^{-3}$\,keV$^{-1}$cm$^{-2}$s$^{-1}$
and $A_p<1.5\times10^{-5}$\,keV$^{-1}$cm$^{-2}$s$^{-1}$). According to
Section~\ref{sec:pulseprofile} the dim phase corresponds to
$\eta_\mathrm{AC1} \approx 5^{\circ}$, a viewing angle almost parallel
to the magnetic field. \citet{nishimura15} specifically addressed
4U\,1626$-$67 and found in simulations that an emission feature could
arise at around 50\,keV for viewing angles nearly parallel to the
magnetic field, which is qualitatively consistent with our
observations as well.

\section{Summary}\label{sec:summary}

We have performed a spectral and timing analysis of the accretion
powered 7.7\,s pulsar 4U\,1626$-$67 during its spin-up phase. The
results are summarized below:

\begin{itemize}

\item The \textit{Fermi}/GBM $\dot{P}$ values during the 2010
  \textit{Suzaku} and 2015 \textit{NuSTAR} observations are different,
  implying a $\sim$15\% increase in the spin-up rate.
\item Comparing the phase-averaged 2010 \textit{Suzaku} and 2015
  \textit{NuSTAR} spectra we found that in addition to the flux
  increase below 20\,keV, the continua differ significantly from each
  other above 25\,keV, with \textit{NuSTAR} data showing less hard
  X-ray flux.
\item Based on the changes in flux, $\dot{P}$, and average spectral
  shape, we suggest that the accretion rate increased between the
  \textit{Suzaku} and \textit{NuSTAR} observations, associated with
  decreasing plasma temperature.
\item Based on the \textit{NuSTAR} data we confirm earlier results
  that the pulse profile is strongly energy dependent and changes from
  being dominated by two narrow peaks in the phase range 0.0--0.5
  below 10\,keV to being dominated by a single broad peak around pulse
  phase 0.75 above 20\,keV.
\item The CRSF around 37\,keV in the phase-averaged and phase 0.5--1.0
  \textit{NuSTAR} spectra could not be adequately modeled with a
  single Gaussian (GABS) or pseudo-Lorentzian (CYAB) optical depth
  profile. A good description was, however, obtained with two GABS
  components, leading to an asymmetric profile that is shallower
  towards lower energies.
\item Possible reasons for an asymmetric CRSF shape are the
  superposition of line profiles from different locations in the
  accretion column or the photon spawning effect.
\item We simultaneously modeled the energy-resolved pulse profiles
using a new relativistic ray tracing code to evaluate emission
patterns. A combination of pencil- and fan-beam emission with a magnetic field nearly aligned with the line of sight reproduces the data well. In this model the narrow double peak in the pulse profile observed below 10\,keV in the first half of the pulse profile is caused by a narrow pencil beam with a small offset to the magnetic field while the flat part of the profile is caused by a fan beam.
Towards higher energies the emission geometry of both components evolves (see Figure~\ref{fig8}). In particular the offset of the pencil beam regarding the magnetic field enlarges, which causes the double peak to move apart regarding phase increasing its contribution in the second half of the pulse profile. 

\item A comparison of our pulse profile modeling with earlier
calculations by \citet{kii86} shows that the deduced changes in emission
pattern may be due to the energy and polarization dependence of the Thomson scattering cross section in a strong magnetic field.

\item  The CRSF parameters obtained for the two characteristic pulse phase ranges
show that the observed CRSF profile depends significantly on the pulse phase.
Moreover, our pulse profile modeling leads us to connect the pulse phase
with the angle of the emitted photons measured with respect to the magnetic
field axis. Connecting these results, we found that the CRSF width decreases
and its depth increases with increasing emission angle. This relation is
expected from theoretical predictions for slab-type geometries
\citep{ise98,sch17b} as well as in asymmetrically-illuminated
cylindrical-geometry models with a magnetic field gradient\citep{nishimura15}.
  
\item We also checked for the possible presence of a CRSF signature in
  emission. Such a feature was tentatively reported for the phase
  0.125--0.250 (dim phase) spectrum of the 2006 \textit{Suzaku}
  observation, i.e., the spin-down phase,
  pre-2008-torque-reversal. Such a feature was only marginally
  detected in the 0.125--0.250 \textit{NuSTAR} spectrum.
\end{itemize}

\acknowledgments

The authors appreciate very much the many constructive comments from the anonymous referee. This work is based on data from the \textit{NuSTAR} mission, a project
led by the California Institute of Technology, managed by the Jet
Propulsion Laboratory, and funded by the National Aeronautics and
Space Administration. We thank the \textit{NuSTAR} Operations,
Software and Calibration teams for support with the execution and
analysis of these observations. This research has made use of the
NuSTAR Data Analysis Software (NuSTARDAS) jointly developed by the ASI
Science Data Center (ASDC, Italy) and the California Institute of
Technology (USA). We also thank the \textit{Suzaku},
\textit{Fermi}/GBM and MAXI team members for their dedicated
support in satellite operations and calibration. This work has been
partially funded by the Deutsche Forschungsgemeinschaft under the DFG
grant number WI 1860$/$11-1. It uses ISIS functions (ISISscripts)
provided by ECAP/Remeis observatory and MIT
(\url{http://www.sternwarte.uni-erlangen.de/isis/}). The figures in
this work have been produced with the \texttt{S-Lang} module
\texttt{slxfig}. KP acknowledges support by NASA's \textsl{NuSTAR}
Cycle 1 Guest Observer Grant NNX15AV17G. MTW is supported by the Chief
of Naval Research. WI is supported by the Special Postdoctoral
Researchers Program in RIKEN and JSPS KAKENHI Grant Number 16K17717.

\bibliography{ref}

\listofchanges

\end{document}